\documentclass{ieeeaccess}

\usepackage[nointlimits]{amsmath}
\usepackage{amsthm}
\usepackage{amssymb, cancel}
\usepackage{graphicx, nicefrac}
\usepackage{color}
\usepackage[dvipsnames]{xcolor}
\usepackage{txfonts}
\usepackage{mathtools}
\usepackage{hyperref}
\usepackage{multirow}
\usepackage{tikz}
\usepackage[T1]{fontenc}
\hypersetup{colorlinks=true, citecolor=blue, linkcolor=blue}
\graphicspath{{./figures/}}

\usepackage{url}
\def\bra#1{{\left\langle #1 \right|}}
\def\ket#1{{\left| #1 \right\rangle}}

\NewSpotColorSpace{PANTONE}
\AddSpotColor{PANTONE} {PANTONE3015C} {PANTONE\SpotSpace 3015\SpotSpace C} {1 0.3 0 0.2}
\SetPageColorSpace{PANTONE}%

\usepackage{cite}
\usepackage{amsmath,amssymb,amsfonts}
\usepackage{algorithmic}
\usepackage{graphicx}
\usepackage{textcomp}
\def\BibTeX{{\rm B\kern-.05em{\sc i\kern-.025em b}\kern-.08em
    T\kern-.1667em\lower.7ex\hbox{E}\kern-.125emX}}
\begin{document}
\history{Date of publication xxxx 00, 0000, date of current version xxxx 00, 0000.}
\doi{10.1109/TQE.2020.DOI}

\title{Learning a quantum computer's capability}
\author{\uppercase{Daniel Hothem}\authorrefmark{1},
\uppercase{Kevin Young}\authorrefmark{1}, \uppercase{Tommie Catanach}\authorrefmark{2}, \uppercase{Timothy Proctor}.\authorrefmark{1}}
\address[1]{Quantum Performance Laboratory, Sandia National Laboratories, Livermore, CA 94550, USA}
\address[2]{Sandia National Laboratories, Livermore, CA 94550, USA}
\tfootnote{This material was funded in part by the U.S. Department of Energy, Office of Science, Office of Advanced Scientific Computing Research, Quantum Testbed Pathfinder Program, and by the Laboratory Directed Research and Development program at Sandia National Laboratories. T.P. acknowledges support from an Office of Advanced Scientific Computing Research Early Career Award. Sandia National Laboratories is a multi-program laboratory managed and operated by National Technology and Engineering Solutions of Sandia, LLC., a wholly owned subsidiary of Honeywell International, Inc., for the U.S. Department of Energy's National Nuclear Security Administration under contract DE-NA-0003525. All statements of fact, opinion or conclusions contained herein are those of the authors and should not be construed as representing the official views or policies of the U.S. Department of Energy, or the U.S. Government.}

\markboth
{Hothem \headeretal: Learning a quantum computer's capability}
{Hothem \headeretal: Learning a quantum computer's capability}

\corresp{Corresponding author: Daniel Hothem (email: dhothem@sandia.gov).}

\begin{abstract}
Accurately predicting a quantum computer's capability---which circuits it can run and how well it can run them---is a foundational goal of quantum characterization and benchmarking. As modern quantum computers become increasingly hard to simulate, we must develop accurate and scalable predictive capability models to help researchers and stakeholders decide which quantum computers to build and use. In this work, we propose a hardware-agnostic method to efficiently construct scalable predictive models of a quantum computer's capability for almost any class of circuits, and demonstrate our method using convolutional neural networks (CNNs). Our CNN-based approach works by efficiently representing a circuit as a three-dimensional tensor and then using a CNN to predict its success rate. Our CNN capability models obtain approximately a $1\%$ average absolute prediction error when modeling processors experiencing both Markovian and non-Markovian stochastic Pauli errors. We also apply our CNNs to model the capabilities of cloud-access quantum computing systems, obtaining moderate prediction accuracy (average absolute error around $2-5\%$), and we highlight the challenges to building better neural network capability models.
\end{abstract}


\titlepgskip=-15pt

\maketitle

\section{Introduction}\label{sec:introduction}
\PARstart{Q}{uantum} computers might offer computational speedups over their classical counterparts on practical problems~\cite{Kim23, Sho97}. But until we have quantum computers with hundreds or thousands of logical qubits, quantum processors will either be too noisy or too small to reliably implement general quantum algorithms~\cite{Su21}. This situation leaves prospective quantum computing users in a proverbial ``no man's land,'' unable to easily determine if a processor is capable of solving their problem. Likewise, as we enter the ``beyond classical'' regime, quantum hardware engineers need new tools to assess, in advance, how their design choices will affect the performance of next-generation systems. To address these issues, we need predictive models of a quantum processor's \emph{capability}~\cite{Proctor2021-wt} that are accurate, scalable, and fast to query.

By a quantum processor's ``capability,'' we mean which quantum circuits a quantum processor can run, and how well it can run them. Capability can be quantified and measured in several ways. Existing methods for testing a quantum processor's capability include algorithmic~\cite{Lub23} and random circuit benchmarks~\cite{Proctor2021-wt,Cro19}. These holistic benchmarks measure a quantum processor's performance on a small number of circuits, then concisely summarize its capability using metrics such as the quantum volume \cite{Cro19} or a volumetric benchmarking plot~\cite{Blu20}. However, holistic benchmarks do not predict how well a processor will execute circuits outside of the benchmark's test suite. 


\Figure[ht!]()[width=.99\linewidth]{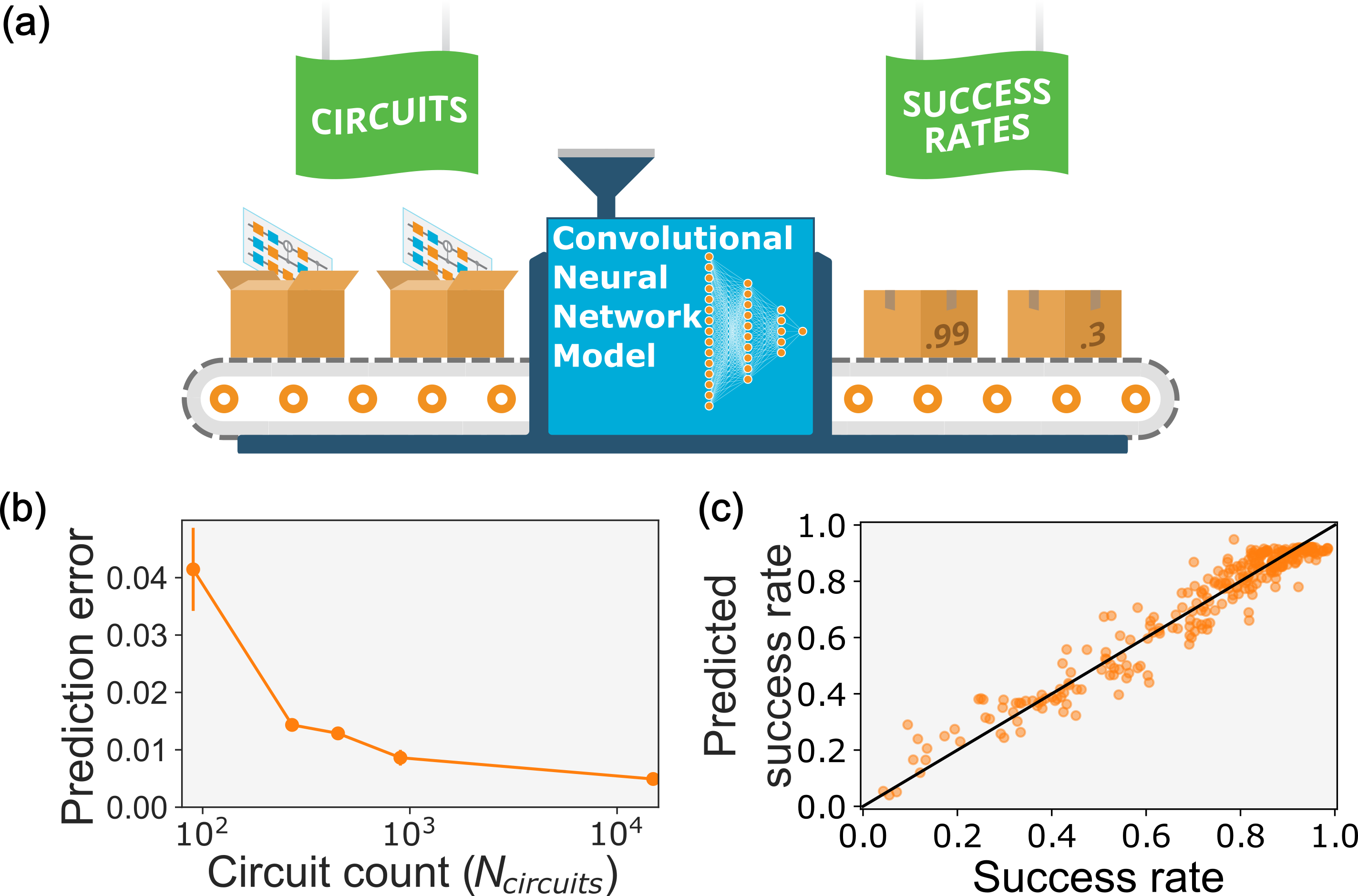}{
\textbf{Quantum capability learning.} \textbf{(a)} We use neural networks to model a quantum computer's capability---a quantification of a quantum computer's performance. Neural network capability models learn to predict how well a quantum processor executes never-before-seen circuits (right side of the figure) from an efficient representations of the circuits (left side of the figure). Predicting which quantum circuits a particular quantum processor can run with low error is essential to choosing which quantum computer to use and which future quantum computers to build. \textbf{(b)} Neural networks can learn capability functions in the presence of Markovian noise (see Section~\ref{sec:5Q-sims}). The decrease in prediction error (y-axis) as training set size grows (x-axis) demonstrates that the neural networks learn increasingly useful features. \textbf{(c)} Neural networks also achieve reasonable prediction accuracy (average absolute error of $2 - 5\%$) when modeling the capabilities of cloud-access quantum computing systems (the plot shows results from \texttt{ibmq\_vigo}). The drop in performance is likely due to the presence of coherent errors (see Fig.~\ref{fig:coherent-errors}).\label{fig:capability-learning}}

In this paper, we show that classical artificial neural networks can approximate a quantum computer's capability (see Fig.~\ref{fig:capability-learning}), and serve as accurate predictive capability models. In particular, we build scalable neural network models for a quantum computer's \emph{capability function} (see Section~\ref{sec:capability}), and demonstrate how these learned models can be used to make predictions of a quantum processor's performance on new circuits that are outside of the set used to train the model. Our method is hardware-agnostic, works for any circuit-based quantum computing platform, and applies to almost any class of circuits (random circuits, benchmarking circuits, algorithmic circuits, etc), although we chose to focus on mirror circuits containing Clifford gates. Our neural network capability models obtain close to a $1\%$ average absolute prediction error when predicting the success rate of quantum circuits executed on a simulated quantum processor experiencing Markovian and non-Markovian Pauli stochastic noise (see Figs.~\ref{fig:markovian-predictions} and~\ref{fig:non-markovian-predictions}). In the latter case, they outperform state-of-the-art Markovian models~\cite{Proctor2021-wt} by nearly an order of magnitude.

Neural network capability models are especially intriguing because they need not be constrained by the assumptions of a particular parameterized error model. This feature has allowed neural networks to be used for many tasks in quantum physics and computing \cite{Kre23, Geb23}---including calibration~\cite{Buk18, Che14, Fos18, Niu19}, circuit compilation~\cite{Mor21, Zhang21}, tomography~\cite{Car19, Sch22, Tor18}, and solving many-body physics problems~\cite{Car17}. As general function approximators~\cite{Hor89}, neural networks can model a processor's capability using a many-parameter ansatz that is highly expressive, potentially enabling them to be trained to model the effect of poorly understood or unexpected error modes (e.g., non-Markovianity) and to operate at scales far beyond those of traditional process matrix-based approaches. 

Here, we show that neural networks can accurately model a processor's capability in the presence of strongly non-Markovian errors---such as gates that get worse over time, or temporal pulse spillover---that conventional error models cannot capture \cite{Nie21}. Furthermore, our neural network capability models are scalable: we show how to efficiently gather training data on a 49-qubit quantum computer, and demonstrate how our trained neural network models can be quickly queried in the many-qubit setting. In principle, our method can scale to hundreds or thousands of qubits. 

The contributions and structure of this paper are as follows. In Section~\ref{sec:capability} we introduce the problem of learning an approximation to a capability function. We conclude in Section~\ref{ssec:pipeline-for-cap-modelling} by outlining a very general, hardware-agnostic pipeline for efficiently constructing scalable capability models of a quantum computer's performance on almost any class of quantum circuits. In Section~\ref{sec:neuralnets} we introduce the convolutional neural network (CNN) architecture and the data encoding that we use in this work. Our representation of the quantum circuits includes a limited form of error sensitivity information that is designed to aid our CNNs in the task of accurately modelling capabilities in the presence of Pauli stochastic errors. This constitutes a simple kind of physics-informed machine learning \cite{Karniadakis2021-yg}. In Sections~\ref{sec:5Q-sims} and~\ref{sec:49Q-sims} we show that CNNs can be trained to accurately model capabilities in the presence of both Markovian and non-Markovian Pauli stochastic errors, including in the many-qubit setting ($n=49$). In Section~\ref{sec:challenges} we highlight some important challenges to creating useful and reliable capability models using neural networks. We demonstrate the difficulties presented by coherent errors, the challenge of generalizing beyond training distributions, and the challenge of extending beyond the limited error sensitivity information included within our data encoding. In Section~\ref{sec:experimental-data} we demonstrate the application of capability learning using CNNs to data from cloud-access quantum computers, obtaining models with moderate prediction accuracy, and we conclude in Section~\ref{sec:discussion}. This paper contains a number of appendices, and in Appendix~\ref{app:prevwork} we discuss related work on circuit fidelity estimation~\cite{Wan22, Vad22, Ame22}.

\Figure[t!]()[width=\linewidth]{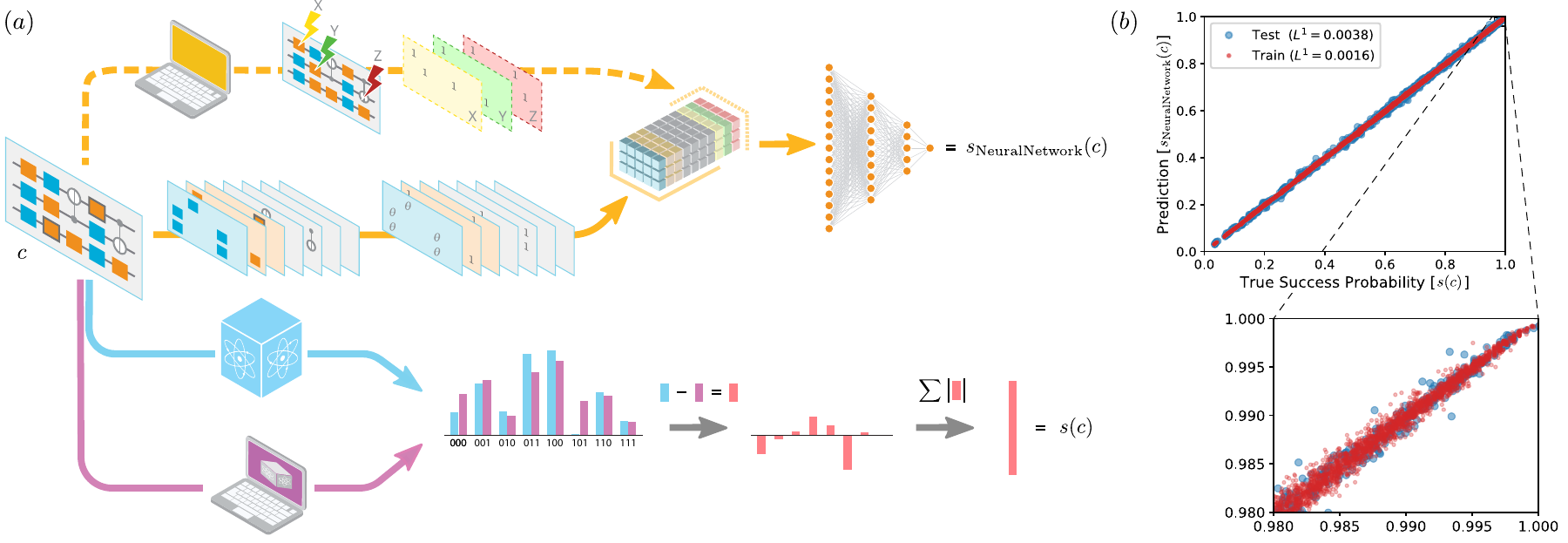}{\textbf{Modelling a quantum computer's capability using neural networks.} \textbf{(a)} Many quantum circuits ($c$) are intended to sample from some distribution (purple histogram), which, in principle, can be calculated by simulating $c$ on a classical computer (purple arrow). However, when $c$ is run on a real quantum processor (blue arrow), hardware errors mean that we sample from a different distribution (blue histogram). The difference between the ideal and actual distributions (red histogram) encodes how well the processor ran $c$, which can be summarized by, e.g., total variation distance (TVD) (red bar). We can quantify how well a processor can run any circuit using a \emph{capability function} ($s$) that maps any circuit $c$ to how well the processor runs $c$ [$s(c)$]. In this work, we aim to construct a model for $s(c)$ using classical artificial neural networks (orange arrows). We input circuits into convolutional neural networks (CNNs) by representing them as three-dimensional tensors that encode which gates are applied in each layer of a circuit (lower orange arrow) and some information about what kinds of errors a circuit is sensitive to (upper orange arrow). In this work, we also define capability functions for circuits that ideally create a quantum state (i.e., they can end with any measurement), or implement a unitary (i.e., they also have an unspecified input), which is not denoted here. \textbf{(b)} The predictions of a CNN trained on random circuits from a hypothetical 5-qubit processor subject to strongly biased local stochastic Pauli errors (see Section~\ref{sec:5Q-sims} for details). This demonstrates that CNNs can accurately model $s(c)$ for a relatively simple but ubiquitous family of errors. To train these CNNs we must gather training data (predictions on training data are shown in red), i.e., a set of circuits each labelled with $s(c)$, and we explain how to do this efficiently in this paper.
\label{fig:schematic}}

\section{Learning a Quantum Computer's Capability}\label{sec:capability}
In this section we introduce the central problem considered in this work: modelling capability functions. The purposes of this section are to (1) introduce the general capability learning problem, and (2) specify the particular form of this problem that we address herein. Sections~\ref{ssec:circuits} and~\ref{ssec:error-models} lay the groundwork for formally defining a quantum computers's capability, which is accomplished in Section~\ref{ssec:capabilityfunc}. Section~\ref{ssec:predictingfid} explains how approximations to a quantum computer's capability can be (inefficiently) learned for nearly any capability metric, and efficiently learned for process fidelity. Section~\ref{ssec:predictingsps} describes the capability learning problem tackled herein---modelling the probability of successful trial of definite-outcome circuits---and relates this problem to the more useful problem of modelling process fidelity. Lastly, Section~\ref{ssec:pipeline-for-cap-modelling} outlines a general framework for building scalable, parameterized models of quantum computer's capability. Only Sections~\ref{ssec:capabilityfunc} and~\ref{ssec:predictingsps} are needed to understand the technical details of this paper, while the remaining sub-sections are useful for contextualizing this paper within the larger problem of understanding a quantum computer's capabilities.




\subsection{Quantum circuits}\label{ssec:circuits}
Here we define what we mean by a quantum circuit, which enables defining capability functions $s(c)$. Quantum circuits are typically defined as a sequence of layers of quantum logic gates. In this work, a $w$-qubit logic layer ($l$) is an instruction to apply physical operations that ideally implements a particular unitary evolution $U(l) \in \text{SU}(2^w)$ on $w$ qubits. This definition excludes logic operations that are intended to be non-unitary, such as mid-circuit measurements. A quantum-input quantum-output (QIQO) $w$-qubit circuit ($c$) over a $w$-qubit logic layer set $\mathbb{L}_w = \{l\}$ is a sequence of $d \geq 0$ layers
\begin{equation}
c = l_d l_{d-1} \cdots l_2 l_1,
\end{equation}
where each $l_i \in \mathbb{L}_w$ \cite{Pro22}. The circuit $c$ is an instruction to apply its constituent logic layers, $l_1$, $l_2$, $\dots$, in sequence, implementing the unitary evolution
 \begin{equation}
U(c) = U(l_d) \cdots U(l_2) U(l_1).
 \end{equation}
Below, it will be convenient to use the superoperator representation of this unitary [$\mathcal{U}(c)$] given by:
 \begin{equation}
     \mathcal{U}(c)[\rho] = U(c)\rho U^{\dagger}(c),
 \end{equation}
 and to denote the perfect action of any circuit $c$ by $\gamma(c)$, i.e., for a QIQO circuit 
 \begin{equation}
     \gamma(c) = \mathcal{U}(c) = \mathcal{U}(l_d) \cdots \mathcal{U}(l_2)\mathcal{U}(l_1).
 \end{equation}

A QIQO circuit can be embedded within other quantum circuits---i.e., any $w$-qubit quantum state can be input into the circuit, and any map can be applied to its output $w$-qubit quantum states. However we may intend to apply a circuit to a fixed input state, followed by a fixed basis measurement. This is important here, because a circuit's intended use impacts the appropriate metric for quantifying how well a processor can run that circuit. A circuit's intended use can be formalized by specifying the intended input space for a circuit as well as the intended set of maps on its outputs. QIQO circuits map quantum states to quantum states. We specify two other important choices for input/output spaces by defining standard-input quantum-output (SIQO) and standard-input classical-output (SICO) circuits \cite{Pro22}. A $w$-qubit SIQO circuit $c = l_d \cdots l_1 l_{\textrm{init}}$ is a QIQO circuit ($l_d\cdots l_1$) with the addition of an initial instruction ($l_{\textrm{init}}$) to initialize each of $w$ qubits in the $\ket{0}$ state. Its perfect action [$\gamma(c)$] creates the pure $w$-qubit quantum state $|\psi(c)\rangle\langle\psi(c)|$ given by:
 \begin{equation}
\gamma(c) = \ket{\psi(c)}\bra{\psi(c)} = \mathcal{U}(l_d\cdots l_1)[ \ket{0}\bra{0}^{\otimes w}].
 \end{equation}
A $w$-qubit SICO circuit $c = l_{\textrm{readout}} l_d \cdots l_1 l_{\textrm{init}}$ is a SIQO circuit ($l_d\cdots l_1l_{\textrm{init}}$) with the addition of a final instruction ($l_{\textrm{readout}}$) to measure all of the qubits in the computational basis. Its perfect action is to draw a sample from a probability distribution $\gamma(c) = \mathsf{P}(c)$ over length-$w$ bit strings $x$, whereby the probability to obtain the bit string $x$ (denoted $\mathsf{P}(c)[x]$) is given by
 \begin{equation}
\mathsf{P}(c)[x] = \textrm{Tr}\left(\ket{x}\bra{x} \mathcal{U}(l_d \cdots l_1) [\ket{0}\bra{0}^{\otimes w}]\right).
 \end{equation}
The three kinds of circuits we consider are summarized in Table~\ref{table:circuits}. All three of these circuit families are part of a broader class of circuits, with mixed quantum-classical inputs and outputs (MIMO), which we do not consider further.

For the purposes of this paper, operations across multiple layers in a circuit must \emph{not} be combined (compiled) together by implementing a physical operation that enacts their composite unitary. This is accomplished by adding in ``barriers'' between circuit layers. These barriers between circuit layers are often used in benchmarking and characterization methods \cite{Mag11,  Proctor2021-wt, Pro22, Nie21}, and it simplifies our prediction task. This is because the inclusion of barriers within circuits removes the need to learn the behaviour of any classical algorithms used to compile together circuit layers \footnote{Note, however, that an individual circuit layer will necessarily be compiled into a processor's native gates. This compilation can also be arbitrarily complex in general, but its complexity is limited in practice if we choose a layer set that closely corresponds to the native circuit layers of the processor in question.}, which can be arbitrarily complex.

\subsection{Modelling an imperfect quantum circuit}\label{ssec:error-models}
The definition of capability functions (below) uses a mathematical model [$\tilde{\gamma}$] for a processor's imperfect implementation of a circuit, and we now introduce this model. Our mathematical model $\tilde{\gamma}$ does not rely on the most widely-used assumptions about a processor's errors (e.g., Markovianity). We avoid encoding those assumptions into our definition of capability functions because methods for learning capabilities have the potential to be accurate even when those assumptions are violated. Our model $\tilde{\gamma}$ assumes that a processor's imperfect implementation of a circuit $c$ depends on $c$ and possibly some auxiliary classical observable ``context'' variable[s] (e.g., time) from some state space $\mathbb{A}$. (No quantum degrees of freedom are allowed.) That is, we use a function $\tilde{\gamma}(c,a)$ to represent the imperfect implementation of $c \in \mathbb{C}$ with context $a \in \mathbb{A}$. Specifically (as summarized in Table~\ref{table:circuits}):
\begin{enumerate}
    \item $\tilde{\gamma}(c,a)$ is an unknown probability distribution over $w$-bit strings [$\tilde{\mathsf{P}}(c,a)$] if $c$ is a SICO circuit,
    \item  $\tilde{\gamma}(c,a)$ is an unknown $w$-qubit quantum state [$\rho(c,a)$] if $c$ is a SIQO circuit, and
    \item $\tilde{\gamma}(c,a)$ is an unknown $w$-qubit completely positive and trace preserving (CPTP) map [$\Lambda(c,a)$] if $c$ is a QIQO circuit.
\end{enumerate}

This framework encompasses all Markovian errors (as defined in Appendix~\ref{app:max-markovian-model} and Ref.~\cite{Nie21}), as well as a wide variety of (but not all) non-Markovian errors. Because $\tilde{\gamma}$ is a general function from circuits to distributions, states, or CPTP maps, this framework can represent the effect of complex error processes within circuits, including: gate errors that increase over the course of a circuit (caused by, e.g., heating in ion-traps); gate error processes that depend on what gates were applied earlier in a circuit (known as serial context dependence); and general crosstalk errors \cite{Sarovar2020-pz}. Because $\tilde{\gamma}$ also depends on observable context variable[s] $a \in \mathbb{A}$, this framework can model the effects of many non-Markovian errors, such as time-varying error processes \cite{Bylander2011-cf, Mavadia2018-ki, Proctor2020-iz, Huo2018-ej} like slow drift (by letting $a$ include wall-clock time, or the time since the last calibration), or the impact of measurable control or environmental parameters. 

\subsection{A quantum computer's capability function}\label{ssec:capabilityfunc}
We now define capability functions, which we aim to learn approximations to. Capability functions are intended to quantify how close a processor's implementation of each circuit $c$, within some circuit set $\mathbb{C}$, is to $c$'s ideal action [$\gamma(c)$]. A capability function is defined using (1) a set $\mathbb{C}$ of circuits, (2) our mathematical model $\tilde{\gamma}$ for a processor's imperfect implementation of a circuit, and (3) a function $\epsilon$ that quantifies the difference between the perfect and imperfect implementations of a circuit. The capability function is a map from circuits ($\mathbb{C}$), and any observable context variables ($\mathbb{A}$) on which $\tilde{\gamma}$ depends, to $\epsilon[\gamma(c), \tilde{\gamma}(c, a)]$. Specifically, the capability function for metric $\epsilon$ is
\begin{equation}
s_{\epsilon}(c, a) =  \epsilon[\gamma(c), \tilde{\gamma}(c, a)].
\end{equation}

There are many well-motivated choices for $\epsilon$, including: the TVD, cross-entropy, or Hellinger (classical) fidelity [for SICO circuits, where $\gamma(c)$ and $\tilde{\gamma}(c,a)$ are probability distributions]; the trace distance or quantum state fidelity [for SIQO circuits, where $\gamma(c)$ and $\tilde{\gamma}(c,a)$ are quantum states]; or the diamond distance \cite{Aharonov1998-ov}, total error \cite{Madzik2022-jh}, or process fidelity \cite{Nielsen2002-iu} [for QIQO circuits, where $\gamma(c)$ and $\tilde{\gamma}(c,a)$ are CPTP maps]. Each metric has a different interpretation---e.g., diamond distance is a form of worst-case error---and the ideal metric with which to define $s_{\epsilon}$ will depend on the intended uses for a model for $s_{\epsilon}$. 

\begin{table}
\begin{center}
\begin{tabular}{ |c|c|c|c| } 
 \hline
Circuit Type & Action &  $\gamma(c)$ &  $\tilde{\gamma}(c,a)$ \\ 
\hline
QIQO & Applies a quantum process & $\mathcal{U}(c)$ &  $\Lambda(c,a)$ \\
SIQO & Creates a quantum state & $\ket{\psi(c)}\bra{\psi(c)}$ &   $\rho(c,a)$  \\
SICO & Samples from a distribution & $\mathsf{P}(c)$ &  $\tilde{\mathsf{P}}(c,a)$ \\
 \hline
\end{tabular}
\end{center}
\caption{A summary of the three types of quantum circuit considered herein, their action (the kind of process they produce), and the mathematical objects and the corresponding notation we use to represent their perfect [$\gamma(c)$] and imperfect [$\tilde{\gamma}(c, a)]$ implementations. See Section~\ref{sec:capability} for the definition of each element in this table.}
\label{table:circuits}
\end{table}

\subsection{Evaluating a capability function}\label{ssec:predictingfid}
To directly learn an approximation to $s_{\epsilon}$ we require labelled training data, i.e., a dataset consisting of a set of circuits $\{c\}$ and (when relevant) contexts $\{a\}$ with each $(c,a)$ paired with an estimate of $s_{\epsilon}(c,a)$. Creating such a dataset requires a method for estimating $s_{\epsilon}(c, a)$. We now discuss whether and how this estimation can be done efficiently. Note that evaluating $s_{\epsilon}(c, a)$ is closely related to the well-known problem of verifying the correctness of the results of a quantum computation.

In principle, a capability function $s_{\epsilon}(c, a)$ can be estimated to any desired precision for any circuit $c \in \mathbb{C}$ and any controllable context $a \in \mathbb{A}$ using a tomographic method. This method consists of (1) running experiments that enable the estimation of $\tilde{\gamma}(c, a)$, and then estimating $\tilde{\gamma}(c, a)$ from the data, (2) computing $\gamma(c)$ by simulating $c$ on a classical computer, and then (3) computing $\epsilon[\gamma(c),\tilde{\gamma}(c,a)]$. For example, if $c$ is a SICO circuit then $\tilde{\gamma}(c, a)$ is the probability distribution that a sample is drawn from in each execution of $c$ in context $a$. In principle, this can be estimated simply by running $c$ many times, in context $a$. Similarly, if $c$ is a QIQO or SIQO circuit, then $\tilde{\gamma}(c)$ is a quantum process or quantum state, respectively, which can be estimated to any desired precision using process or state tomography \cite{Nie21}, respectively (or, to avoid known inconsistencies in state and process tomography, using gate set tomography (GST) \cite{Nie21}). This procedure consists of embedding $c$ within a variety of circuits and then inferring $\tilde{\gamma}(c, a)$ from the data. However, for a general circuit $c$, the tomographic method for evaluating $s_{\epsilon}(c, a)$ is well-known to be inefficient. In general, this method for evaluating $s_{\epsilon}(c, a)$ requires classical computations (to simulate $c$) that are exponentially expensive in the number of qubits ($n$) on which $c$ acts, and a number of circuit executions that is also exponentially large in $n$. So capability learning based on a tomographic approach to estimating $s_{\epsilon}(c, a)$ is inefficient, in general.

Direct and efficient methods for estimating the value of a capability function $s_{\epsilon}(c,a)$ are critical for direct learning of capability functions. The circumstances under which $s_{\epsilon}(c,a)$ can be efficiently estimated is an interesting open question, i.e., for which $\epsilon$ and under what assumptions about a processor's errors [which includes assumptions about $\tilde{\gamma}(c,a)$] is there an efficient method for estimating $s_{\epsilon}(c,a)$? However, there is a well-motivated choice for $\epsilon$ for which an efficient method for estimating $s_{\epsilon}(c,a)$ is known: process fidelity\footnote{Process fidelity comes in two variants: average gate fidelity and entanglement fidelity, that are linearly related by a dimensionality factor \cite{Nielsen2002-iu}, and herein we use entanglement fidelity.}. Process fidelity ($F$) is defined as
\begin{equation}
    F[\gamma(c), \tilde{\gamma}(c,a)] = \bra{\varphi} \gamma^{\dagger}(c) \tilde{\gamma}(c,a)[\ket{\varphi}\bra{\varphi}] \ket{\varphi},
\end{equation}
where $\varphi$ is any maximally entangled state in a doubled Hilbert space \cite{Nielsen2002-iu}. Almost any circuit's process fidelity can be efficiently estimated using mirror circuit fidelity estimation (MCFE) \cite{Proctor2022-es} (under certain assumptions about the underlying error processes, detailed in Ref.~\cite{Proctor2022-es}). Because a capability function defined by process fidelity can be evaluated using a method that is efficient in the number of qubits $n$, it is feasible that an approximation to $s_F$ can be learned even in the many-qubit setting, where approximate capability function models will be most useful.

\subsection{The capability function for definite outcome circuits}\label{ssec:predictingsps}
The capability function defined by process fidelity ($s_F$) is well-motivated, and learning an approximation to $s_F$ is feasible, in the sense that training data can be efficiently obtained. However, in this work we apply neural networks to a different but related problem. Instead we consider the problem of learning a capability function for \emph{definite outcome circuits} (defined below), which are a subclass of SICO circuits---that is, we consider $s$ defined over a circuit set $\mathbb{C}$ containing only definite outcome circuits. In Appendix~\ref{app:spl} we explain why we choose to address this problem, rather than process fidelity learning, and why we conjecture that a neural network method that can accurately model $s$ for definite outcome circuits will, with minor adaptions, be able to accurately model $s_F$ when trained on circuit process fidelities.

A SICO circuit $c$ is a definite outcome circuit if and only if its error-free output distribution $\gamma(c)$ has support only on a single ``success'' bit string $x_{\textrm{s}}(c)$, i.e.,  
\begin{equation}
\gamma(c)[x_{\textrm{s}}(c)] = 1.
\end{equation}
For definite outcome circuits, the probability that a circuit $c$ outputs its success bit string $x_{\textrm{s}}(c)$ is the single natural choice for $\epsilon$ \footnote{All standard choices for $\epsilon$ with SICO circuits are, when applied only to definite outcome circuits, equivalent to the probability of the correct bit string. This includes the TVD and the Hellinger fidelity.}. Therefore, the unique well-motivated definition for the capability function of definite outcome circuits is simply
\begin{equation}
s(c, a) =\tilde{\gamma}(c, a)[x_{\textrm{s}}(c)],
\end{equation}
which is the circuit's ``success probability''. The success probability of a definite outcome circuit $c$ can be efficiently estimated whenever the success bit string $x_{\textrm{s}}(c)$ can be efficiently computed on a classical computer. In particular we can estimate $s(c, a)$ [denoted $\hat{s}(c, a)$] from $N_{\rm shots}$ executions of $c$ in context $a$ as
\begin{equation}
\hat{s}(c, a) = \frac{N_{x_{\textrm{s}}(c)}}{N_{\rm shots}} \label{eq:estsc}
\end{equation}
where $N_{x_{\rm{s}}(c)}$ is the number of times the bit string $x_{\rm{s}}(c)$ is output from the circuit. 

Almost any circuit $c$ can be turned into a definite outcome ``mirror circuit'' $m(c)$ \cite{Pro22, Proctor2021-wt, Hines2022-xv, Mayer2021-vl} for which $x_{\textrm{s}}[m(c)]$ can be efficiently computed, and all the circuit sets used in our numerical experiments---i.e., the applications of our CNNs to simulated or experimental data---contain only mirror circuits $m(c)$. Circuit mirroring is a motion-reversal circuit or Loschmidt echo (i.e., following $c$ by its inverse) that is modified to prevent the systematic cancellation of errors that can occur within standard motion-reversal circuits.

\subsection{A general pipeline for constructing neural network capability models}\label{ssec:pipeline-for-cap-modelling}
We now lay out a general, hardware-agonistic pipeline for efficiently constructing scalable, parametrized models of $s_F$ on almost any class class of circuits. The steps are:
\begin{enumerate}
    \item gather a representative sample of circuits $C \subset \mathbb{C}$,
    \item use MCFE to efficiently estimate $s_F(c)$ for every circuit in $C$,
    \item select a parameterized model $\mathcal{M}$,
    \item encode each circuit $c\in C$ in a format $E(c)$ that is processable by $\mathcal{M}$,
    \item use the tuples $(E(c), s_F(c))$ to fit the model $\mathcal{M}$.
\end{enumerate}
This pipeline is efficient and scalable in the number of qubits whenever the number of parameters in $\mathcal{M}$ and the encoding process scale polynomially in $n$. We spend the remainder of the paper demonstrating that both conditions are met when $\mathcal{M}$ is a CNN, and we explore the advantages and disadvantages of using a neural network.

\section{Predicting capabilities using convolutional neural networks}\label{sec:neuralnets}
In this section we introduce our method for predicting circuit success probabilities using CNNs. In this work we aim to predict the success probabilities [s(c)] of definite outcome circuits, run on a specific quantum processor, and we consider no context information (i.e., $\mathbb{A}$ from Section~\ref{sec:capability} is trivial). So, the input of our neural networks is a definite outcome circuit $c$, and the output is a prediction for the success probability $s(c)$ that would be observed if $c$ were run on the processor that we are modelling. To address this prediction problem using neural networks we must choose: (1) the set of quantum circuits whose success probabilities we aim to predict (see Section~\ref{ssec:circuit-generation}); (2) a representation for the circuits that enables inputting them into a neural network (see Section~\ref{ssec:encoding}); (3) a structure for the neural networks (see Section~\ref{ssec:cnns}); (4) methods for training the parameters [i.e., weights] of the neural network (see Section~\ref{ssec:training}, and tuning its hyperparameters (see Section~\ref{ssec:tuning}); and (5) methods for evaluating the final model's performance (see Section~\ref{ssec:metrics}).

\subsection{Circuit selection}\label{ssec:circuit-generation}
Training, hyperparameter tuning, and evaluation of our neural networks requires training, validation, and test datasets. These datasets each consist of circuits $\{c\}$ that are each labelled with an estimate [$\hat{s}(c)$] of that circuit's success probability [$s(c)$]. We denote these datasets by $D_{\textrm{train}}$, $D_{\textrm{validate}}$, and $D_{\textrm{test}}$ herein. Here we explain how we select the training, validation, and test circuit sets (in Sections~\ref{ssec:predictingsps}-\ref{ssec:predictingfid} we explained how we estimate $s(c)$ for any circuit $c$). Each of these circuit sets is sampled from a set $\mathbb{C}$ defining the set of all possible circuits (the set $\mathbb{C}$ used in our examples is introduced below).

We construct training, validation, and (in-distribution) test circuit sets using a parameterized distribution $p(\alpha): \mathbb{C} \rightarrow \mathbb{R}$. First a circuit set $D_{\textrm{combined}}$ is constructed by (1) either systematically varying the distribution's parameters $\alpha$ (corresponding, e.g., to the circuits' depths) or randomly sampling the distributions parameters, and (2) drawing independent samples from $p(\alpha)$ for each selected parameter value $\alpha$. The sampled circuit set $D_{\textrm{combined}}$ is then randomly partitioned into training, validation, and (in-distribution) test circuit sets. This is consistent with standard practices in machine learning.

Evaluating a model's prediction accuracy on test circuits drawn from the same distribution as the training circuits corresponds to standard practice, and neural network models often do not generalize well to data drawn from a different distribution. However, the utility of a neural network model for $s(c)$ will depend on its ability to accurately predict the performance of those circuits that are of most interest to a user of this model for $s(c)$ (e.g., perhaps only circuits that implement a particular algorithm are of interest), and the relevant circuit set[s] might not be known at the time of model training (therefore preventing the relevant circuit set[s] from being used to define $p$). In Sections~\ref{ssec:ood-wide-predictions} we explore whether CNN models for $s(c)$ generalize to additional test data ($D_{\text{test}}$), containing circuits drawn from a distribution $p'$ that differs from $p$. This is an example of what is known as \emph{out-of-distribution} prediction or generalization. 

Specific circuit sets are sampled from the set of all possible circuits ($\mathbb{C}$) and we now specify how $\mathbb{C}$ is defined in this work. We consider circuit sets for an $n$-qubit processor (we label the qubits by $\mathbb{Q}= \{1,\dots,n\}$) with a set of logic layers that is specified by a directed connectivity graph ($G$) over the qubits $\mathbb{Q}$, a two-qubit gate set $\mathbb{G}_2$, and a one-qubit gate set $\mathbb{G}_1$. We consider all $w \leq n$ qubit (SICO) circuits, over a $w$-qubit subset of $\mathbb{Q}$ ($\mathbb{Q}_w$), consisting of layers that contain parallel applications of gates from $\mathbb{G}_1$ and two-qubit gates from $\mathbb{G}_2$ that respect the processor's connectivity graph. Our circuit encoding (see Section~\ref{ssec:encoding}) assumes a connectivity graph that can be embedded in a square grid (however, extensions to other connectivity graphs are simple). Our circuit encoding assumes a single two-qubit gate, and in all our numerical examples
\begin{equation}
    \mathbb{G}_2 = \{\text{CNOT}\}.
\end{equation}
The circuit encoding assumes a single-qubit gate set in which each gate can (but need not be) parameterized by a continuous variable. For all our simulated datasets
\begin{equation}
\mathbb{G}_1 = \{ Z(\theta), X_{\nicefrac{\pi}{2}}, X_{-\nicefrac{\pi}{2}} \},\label{eq:G1}
\end{equation}
where $Z(\theta)$ is a $Z$ rotation by $\theta$, i.e., $U[Z(\theta)] = \exp(-i\theta Z /2)$, and $X_{\nicefrac{\pi}{2}}$ and $X_{-\nicefrac{\pi}{2}}$ are $X$ rotations by $\nicefrac{\pi}{2}$ and $-\nicefrac{\pi}{2}$, respectively. For all datasets from cloud-access quantum computers
\begin{equation}
\mathbb{G}_1 = \mathbb{C}_1,\label{eq:G1-exp}
\end{equation}
where $\mathbb{C}_1$ is the set of 24 single-qubit Clifford gates. Finally, our current circuit encoding requires that every gate in every circuit is a Clifford gate [so, for the gate set of Eq.~\eqref{eq:G1}, this means $\theta \in \{-\nicefrac{\pi}{2}, 0, \nicefrac{\pi}{2}, \pi\}$]. The restriction to Clifford circuits is necessary for an \emph{optional} part of our circuit encoding that we conjecture can be adapted to general circuits (see the discussion in Section~\ref{ssec:encoding} and Section~\ref{ssec:no-stabilizers}).

Throughout this paper our training, validation and test circuit sets are sampled from two families of parameterized distributions over circuits: randomized mirror circuit \cite{Pro22, Proctor2021-wt, Hines2022-xv, Mayer2021-vl} and periodic mirror circuits \cite{Proctor2021-wt} (see Fig.~1 of Ref.~\cite{Proctor2021-wt} for a diagrammatic representation of these circuits, and the supplemental material therein for comprehensive definitions). Depth $d$ randomized mirror circuits consist of $\nicefrac{d}{2}$ independently sampled layers of gates, followed by $\nicefrac{d}{2}$ layers consisting of the inverse of that circuit (with some added randomization to prevent systematic error cancellation). In contrast, periodic mirror circuits consist of repeating the same short (randomly sampled) sequence of gates many times, followed by the inverse circuit (again with some added randomization to prevent systematic error cancellation). Randomized mirror circuits are highly disordered---and they are similar in nature to the random circuits used in many benchmarking methods (e.g., \cite{Aru19, Mag11, Cro19})---whereas periodic mirror circuits can amplify coherent errors. Each distribution is parameterized by: (1) circuit width ($w$); (2) circuit depth ($d$); (3) the subset $\mathbb{Q}_w \subset \{1,\dots,n\}$ of $w$ connected qubits that the circuit runs on; and (4) expected two-qubit gate density ($\xi$). We aim to model $s(c)$ for the variables $w$, $d$ and $\mathbb{Q}_w$, so we either systematically vary these three parameters or sample them randomly.

\subsection{Circuit encoding}\label{ssec:encoding}
To learn an approximation to a capability function $s(c)$ we must choose a mathematical representation $I(c)$ for the circuits that can be input into the chosen neural network architecture. The neural network is tasked with approximating $s(c)$ given $I(c)$, so the complexity of its learning task depends on the choice for this representation. The learning problem is easier if we use a representation of the circuits that makes it easy for the chosen neural network architecture to extract features of circuits that are highly predictive of $s(c)$ \footnote{To see this, note that the learning problem is infeasible if we used an encoding that encrypts the circuits}. We represent a $w \times d$ circuit for an $n$-qubit processor (where $w \leq n$) using a $n \times d$ image with multiple ``color channels'', as illustrated in Fig.~\ref{fig:schematic}. That is, we represent the circuit $c$ by an $n \times d \times h$ tensor $I(c)$ where $h$ is the number of channels. The $(i,j)$ ``pixel'' of the image [$I_{ij}(c)$], meaning the vector
\begin{equation}
    I_{ij}(c) = \left(I_{ij1}[c], I_{ij2}[c],\dots, I_{ijh}[c]\right)^T,
\end{equation}
stores information about what happens to qubit $i$ in layer $j$ of the circuit. So this encoding preserves the locality of consecutive layers in the circuit. However, it does \emph{not} encode information about the spatial arrangement of the physical qubits (our labelling of the qubits is arbitrary). The channels are split into two kinds of channel: gate channels and error sensitivity channels, introduced in turn below.

The gate channels are used to encode which gate is being applied to each qubit in each layer, using a modified one-hot encoding. Our encoding uses $h_{\rm gate} = |\mathbb{G}_1| + 4$ gate channels: one for each single-qubit gate in $\mathbb{G}_1$ and four channels to encode CNOT gates \footnote{When we applied our techniques to data from cloud-access quantum computers we use a slightly different encoding, which can be found in the supplemental data and code.}. We use four channels for the CNOT gates as each qubit has at most four neighbours in the connectivity graph, so four channels is sufficient for a lossless encoding of the CNOT gates. The channels correspond to a CNOT gate with a neighbour that is to the left, to the right, above, or below the qubit in question [which is qubit $i$ for pixel $I_{ij}(c)$]. If in layer $j$ of circuit $c$ the gate $G$ is applied to qubit $i$ and $G$ is encoded in channel $k$ then $I_{ijk}(c)= v(G)$, with the value in all other channels set to zero (as in one-hot encoding). If $G$ is a single-qubit parameterized gate then $v(G)=\theta$ where $\theta$ is the gate's parameter value [so, e.g., $v(G)=\theta$ for a $Z(\theta)$ gate], and if $G$ is a single-qubit gate with no parameters then $v(G)=1$. If $G$ is a CNOT gate then $v(G)=1$ ($v(G)=-1$) if qubit $i$ is the control (target) qubit. Therefore, the value $v(G)$ stored in a channel includes any information about the identity of the gate that is being applied that is not encoded into the channel index.

Machine learning techniques applied to physics problem are often more accurate if known physics is encoded into the methods \cite{Karniadakis2021-yg} (known as physics-informed machine learning). We implement a simple form of physics-informed machine learning by encoding into $I(c)$ some information about what errors the circuit $c$ is sensitive to. This is the role of our \emph{error sensitivity channels}. The error sensitivity channels are used to encode, into pixel $I_{ij}(c)$, information about which kinds of errors qubit $i$ is sensitive to when layer $j$ of $c$ is applied. In our encoding, we utilize three error sensitivity channels $h_X$, $h_Y$ and $h_Z$. For $P \in \lbrace X, Y, Z\rbrace$, $h_P$ encodes whether the single-qubit Pauli error $P$ on qubit $i$ at layer $j$ would transform the qubits into an orthogonal state when applied to the ideal state of the system after layer $j$. Specifically, letting
\begin{equation}
|\psi_j\rangle = U(l_j)\cdots U(l_1)|0\rangle^{\otimes n},
\end{equation}
for the circuit $c=l_d\dots l_1$, then 
\begin{equation}
    I(c)_{ijh_P} = 1 - |\langle\psi_j\vert P_i \vert\psi_j\rangle|.
\end{equation}
Thus, $I(c)_{ijh_P} = 0$ if $|\psi_j\rangle$ is an eigenstate of $P_i$ and otherwise $I(c)_{ijh_P} = 1$. Here $P_i$ denotes Pauli $P$ operator on qubit $i$ (tensored with an identity on all other qubits). 

A complete description of $\psi_j$ could be encoded into $I(c)$ using $O(n)$ bits (at each pixel) that specify a set of generators for $\psi_j$'s stabilizer group. This is because $\psi_j$ is a stabilizer state, as we consider only Clifford circuits herein. We do not do so for two reasons. First, our encoding provides a CNN with easy access to information about the impact of a particular important kind of errors---local stochastic Pauli errors---which is not easily extracted from an arbitrarily chosen set of generators for $\psi_j$'s stabilizer group. Second, we conjecture that limited error sensitivity information similar to that encoded here can be obtained even for general, non-Clifford circuits, whereas encoding a complete but efficient description of the state $\psi_j$ does not generalize to arbitrary non-Clifford circuits.

\Figure[t!]()[width=\linewidth]{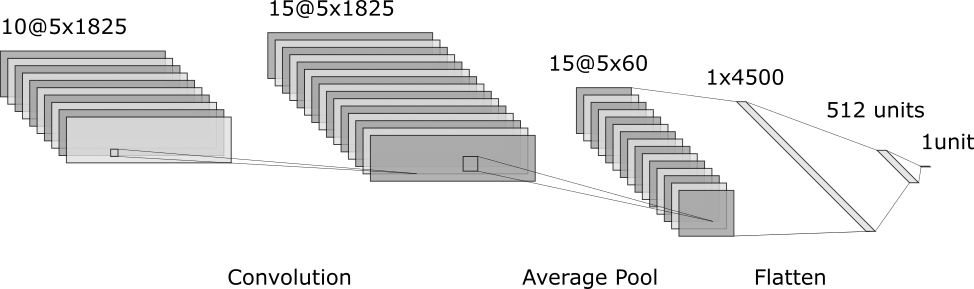}{\textbf{A convolutional neural network for predicting circuit success probabilities.} In this work we explore using CNNs to predict a circuit $c$'s success probability $s(c)$ when run on a particular quantum computer. Here we provide an example of the CNN architecture that we use for modelling $s(c)$. The input to our CNNs is an image representation $I(c)$ of a circuit $c$, where $I(c) \in \mathbb{R}^{n \times d \times h}$ and $n$ is the number of qubits ($c$'s width), $d$ is $c$'s depth, and $h$ is the number of channels used in the encoding. In this example, $n=5$, $d=1825$ and $h=10$. Our CNNs consist of one or more layers of convolutional filters (here there is one convolutional layer, with 15 kernels of shape $(1,4)$), interspersed with dimension-reducing average pooling layers (here there is one pooling layer, which averages across the depth dimension), followed by a multi-layer perceptron consisting of dense layers of neurons (here there are two dense layers). The final dense layer contains a single neuron with a sigmoid activation function, so that the output---the model's prediction for $s(c)$---is within $[0,1]$. Convolutional and dense layers contain parameters (weights and biases) that are learned in the training process: the example shown here contains 2,305,640 parameters. The structure of the CNN is selected using hyperparameter tuning. The example shown here is the CNN used for the local Pauli stochastic error model trained on $N_{\text{circuits}} = 14940$ and $N_{\text{shots}} = \infty$ (see Fig.~\ref{fig:markovian-predictions}). The image was generated using \cite{LeN19}.
\label{fig:nm-architecture}}

\subsection{Convolutional neural networks}\label{ssec:cnns}

There are many neural network architectures that could be applied to the problem of capability modelling. In this work we use CNNs \cite{Osh15, Col08, Goo16}. CNNs extract predictive features from their input using learned convolutional filters, and they have proven useful for image classification~\cite{Osh15} and natural language processing~\cite{Col08}. CNNs are a promising architecture for capability modelling because convolutional filters can pick out particular patterns of gates in a circuit (encoded as an image, or tensor), and particular but \emph{a priori} unknown patterns of gates can be correlated with $s(c)$ \cite{Sarovar2020-pz, Pro22}. Furthermore, it is possible to construct CNNs whose complexity---i.e., the number of parameters that must be learned---increases only slowly (or even not at all) with the number of qubits ($n$) in a processor, meaning that scalable capability modelling with CNNs is feasible. An example of the CNNs we used is shown schematically in Fig.~\ref{fig:nm-architecture}, while full technical details are given in Appendix~\ref{app:cnns}.


See Goodfellow \emph{et al.}~\cite{Goo16} or the Appendix for a more detailed introduction to CNNs.


\subsection{Network training}\label{ssec:training}
CNNs are trained to approximate a function by iteratively modifying all of their learnable parameters, to improve the model's predictions on training data. We optimize a network's weights using the Adam optimization algorithm, a gradient-based optimization method \cite{Kin14}. A round of training consists of: (1) evaluating the network's predictions on $D_{\text{train}}$ using a loss function and (2) updating the network's weights to minimize the training loss. This process is repeated for some number of \emph{epochs} (the number of training epochs is a hyperparameter, as discussed below).

We use the average binary cross-entropy (BCE) as our loss function. The average BCE of a model $\mathcal{M}$'s predictions $\boldsymbol{s}_{\mathcal{M}} =\{s_{\mathcal{M}}(c)\}$ to a set of observations $\hat{\boldsymbol{s}} =\{\hat{s}(c)\}$ is
    \begin{align*}
        H(\hat{\boldsymbol{s}},\boldsymbol{s}_{\mathcal{M}}) &= - \frac{1}{N_{\textrm{circuits}}}\sum_{c}s_{\mathcal{M}}(c)\log[\hat{s}(c)],
    \end{align*}
where $N_{\textrm{circuits}}$ is the number of circuits in the dataset. A circuit's success probability $s(c)$ is estimated using $\hat{s}(c) = N_{x_{\textrm{s}(c)}}/N_{\rm shots}$ [see Eq.~\eqref{eq:estsc}] where $N_{\rm shots}$ is the number of times each circuit is repeated. Whenever $N_{\rm shots}$ is finite and equal for all the circuits in a dataset then $H(\hat{\boldsymbol{s}},\boldsymbol{s}_{\mathcal{M}})$ is equal to the log-likelihood of the data given the model's predictions multiplied by a multiplicative factor (of $-N_{\rm shots}N_{\rm circuits}$). These conditions are satisfied for many of our datasets, and in those cases minimizing the average BCE is equivalent to maximizing the likelihood of the model given the data.

\subsection{Hyperparameter tuning}\label{ssec:tuning}
A CNN's weights and biases are optimized when training the network, but there are many other parameters that can also affect a CNN's prediction accuracy. Such parameters are called \emph{hyperparameters}. Hyperparameters  include (but are not limited to): (i) the number of convolutional, pooling, and dense layers, (ii) the number of neurons in each dense layer, (iii) the shape of each kernel in the convolutional layers, and (iv) the number of training epochs.  Optimal values for hyperparameters are searched for by hyperparameter tuning.

Hyperparameter tuning consists of searching over a space of candidate values for the hyperparameters (we used Bayesian optimization \cite{Moc89, Sno12}). At each step in the search process, a CNN is created with the candidate hyperparameter values and trained using the training dataset ($D_{\text{train}}$). Each model is evaluated on the validation dataset ($D_{\text{validate}}$), and the hyperparameters with the smallest loss on $D_{\text{validate}}$ are chosen. A separate validation set is used because it mitigates the effects of over-fitting each CNN's weights to the training data. After hyperparameter tuning is complete, a new CNN is created using the hyperparameters associated with the lowest loss on $D_{\text{validate}}$. This network is trained on $D_{\text{train}}\cup D_{\text{validate}}$ (this is standard practice in machine learning). For expediency, we often refer to the combined training and validation dataset as the ``training'' data when no confusion will arise. 

The hyperparameter spaces used in our numerical experiments are provided in the supplementary data and code. Our hyperparameter optimizations included varying the shape of the convolutional kernels, as different kernel shapes $(k_w,k_d)$ can create feature maps that extract circuit features that are relevant for different kinds of error. We varied the width of the kernels, as width-$k_w$ kernels jointly analyze the gates applied to $k_w$ qubits, so they can extract circuit features that are relevant for modelling the effects of $k_w$-qubit crosstalk. As our encoding does not preserve the spatial locality of qubits, for few qubit processors (small $n$) we include kernels of width up to $k_w =n$. We varied the length ($k_d$) of the kernels, as length-$k_d$ kernels jointly analyze $k_d$ circuit layers, so they can extract circuit features that are relevant for modelling the effects of errors that depend on $k_d$ sequential layers (e.g., serial context dependence).

\subsection{Evaluating model performance}\label{ssec:metrics}
To evaluate the performance of a model we quantify its prediction accuracy on one or more test datasets ($D_{\text{test}}$), which were not used during training or hyperparameter tuning. We used three complimentary figures of merit to evaluate the performance of a model on test data: Kullback-Leibler (KL) divergence,  the mean absolute error ($L^1$ error), and the Pearson correlation coefficient ($r$). KL divergence is defined by
\begin{align*}
d_{\text{KL}}(\hat{\boldsymbol{s}},\boldsymbol{s}_{\mathcal{M}}) &= H(\hat{\boldsymbol{s}},\boldsymbol{s}_{\mathcal{M}}) - H(\hat{\boldsymbol{s}}),
\end{align*}
where $H(\hat{\boldsymbol{s}}) = H(\hat{\boldsymbol{s}}, \hat{\boldsymbol{s}})$ is the entropy of $\hat{\boldsymbol{s}}$. We use KL divergence as a figure of merit in part because the mean $H(\hat{\boldsymbol{s}},\boldsymbol{s}_{\mathcal{M}})$ is the loss function in the training. The KL divergence removes the entropy of the dataset from $H(\hat{\boldsymbol{s}},\boldsymbol{s}_{\mathcal{M}})$, facilitating easier comparisons between a model's performance on datasets with different entropies. The mean absolute error (or $L^1$ error) defined by 
\begin{align*}
d_{L^1}(\hat{\boldsymbol{s}},\boldsymbol{s}_{\mathcal{M}}) =\frac{1}{N_{\textrm{circuits}}}\sum_{c} | s_{\mathcal{M}}(c) - \hat{s}(c)|,
\end{align*}
 and the Pearson correlation coefficient ($r$) were chosen due to their straightforward interpretations, and to allow comparison to other work. Note, however, that $r=1$ does not imply perfect prediction accuracy. This is because $r$ quantifies the linear correlation between a model's predictions and the data.
 
 \subsection{Predicting capabilities using error rates models}\label{ssec:param-models}
Neural network models for $s(c)$ will be useful if their predictions are sufficiently accurate. A particular task may require a model [$s_{\mathcal{M}}(c)$] for $s(c)$ that achieves a certain accuracy threshold (e.g., $1\%$ or less absolute error on every circuit within some circuit set), and whether a particular neural network model for $s(c)$ satisfies such a criteria can be judged given that task. But a neural network model for $s(c)$ is also only useful if its prediction accuracy is at least as good as other available and equally convenient (e.g., as fast to query) models for $s(c)$. This can be assessed without a particular use-case for $s_{\mathcal{M}}(c)$. Herein, we compare the predictions of our CNNs to that of an \emph{error rates model} (ERM) \cite{Proctor2021-wt}, which we now introduce.

An ERM is a parameterized error model for a processor that consists of modelling each of a processor's logic operation by an error rate ($\epsilon$). The model's prediction for $s(c)$ approximately corresponds to multiplying together the success rates ($1-\epsilon$) for every logic operation in $c$. An ERM's parameters consist of one-qubit gate error rates $\{\epsilon(G,i)\}_{i \in \mathbb{Q}, G \in \mathbb{G}_1}$, two-qubit gate error rates $\{\epsilon(G,i,j)\}_{(i,j) \in \mathbb{E}, G \in \mathbb{G}_2}$ where $\mathbb{E}$ is the set of all connected pairs of qubits, and readout error rates  $\{\epsilon(i)\}_{i \in \mathbb{Q}}$. An ERM's prediction for $s(c)$ is approximately given by 
\begin{equation}\label{eqn:erm_approx}
    s(c) = \prod_{i \in \mathbb{Q}_w} (1-\epsilon(i)) \prod_{g\in c}(1-\epsilon(g)), 
\end{equation}
where $\mathbb{Q}_w$ is the set of qubits on which $c$ acts, and $g \in c$ runs over all the gates (labelled with the qubits on which they act) in $c$. The exact formula for predicting $s(c)$ from an ERM is obtained by using the error rates $\{\epsilon\}$ to construct a global depolarization model \footnote{Like most widely-used parameterized error models, ERMs can be constructed by placing restrictions on the maximal Markovian model (see Ref.~\cite{Nie21} or Appendix~\ref{app:max-markovian-model}), which models a processor's operations using general $n$-qubit CPTP maps.} [and it differs from Eq.~\eqref{eqn:erm_approx} only by $O(\nicefrac{1}{2^n})$ factors]. That formula can be found in Ref.~\cite{Proctor2021-wt} (see Eqs.~(54)-(55) in the supplemental material of Ref.~\cite{Proctor2021-wt}).

ERMs are useful models against which to compare our CNNs because an ERM's parameters can be efficiently estimated from data (for any number of qubits $n$), ERMs are fast to query, and (unlike CNNs) ERMs have interpretable parameters. ERMs have these properties because (1) they contain at most $O(n^2)$ parameters that must be learned from data [and, if the qubit's connectivity is a planar graph then an ERM contains only $O(n)$ parameters], and (2) the prediction for $s(c)$ can be quickly evaluated for any circuit $c$. For these reasons, an ERM is arguably preferable to a neural network model for $s(c)$ if the two models have equal prediction accuracy. Throughout this work, we fit an ERM to the same data used to train a neural network (we use maximum likelihood estimation to fit ERMs). 

\Figure[t!]()[width=\linewidth]{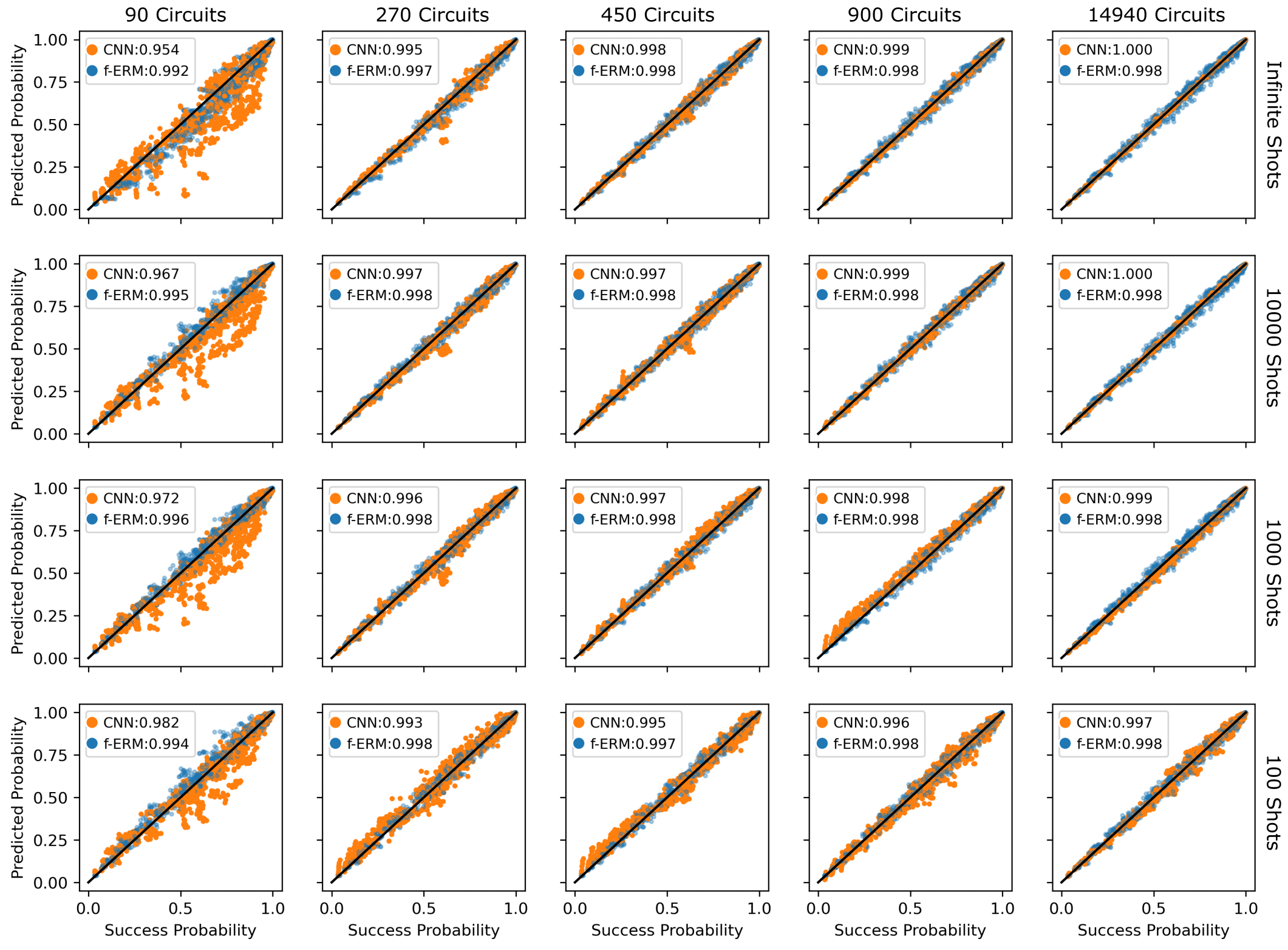}
{\textbf{Modelling the effect of stochastic errors with CNNs.} The prediction accuracy of CNNs trained on simulated data from few-qubit random circuits with a stochastic Pauli errors model. We constructed a circuit set consisting of 16600 randomized mirror circuits on 1-5 qubits, split it into training, validation and testing circuits (a 70\%, 20\%, 10\% split), simulated each circuit under a single biased stochastic Pauli errors model to compute $s(c)$, and then trained and tuned CNNs on sub-sampled datasets of varying quality---constructed by independently varying the training dataset size ($N_{\textrm{circuits}}$) and the shot count ($N_{\textrm{shots}}$), i.e., the number of repetitions of each circuit $c$ used to estimate $s(c)$. We also fit an ERM (error rates model) to the same training data, for comparison. Each subplot shows the true success probabilities $s(c)$ versus the predictions of the CNN (orange) and the fit ERM (f-ERM, blue) evaluated on the full set of 1660 test circuits, for a single (nested) training dataset instance at each (combined validation and) training dataset size and shot count. We observe that the CNN's predictions improve as the dataset size increases and shot noise decreases. Legends show each model's correlation coefficient $r$ (note that $r=1$ implies perfect linear correlation, not perfect predictions). The improvement in the CNN's accuracy with improving dataset quality is quantified in Fig.~\ref{fig:markovian-heatmaps}.
\label{fig:markovian-predictions}}

\section{Predicting capabilities with few qubits and stochastic errors}\label{sec:5Q-sims}

Accurate modelling of capability functions for real quantum processors using neural networks will require an architecture that can (1) model the effects of common kinds of errors, and (2) be trained with feasible amounts of data. In this and the following two sections, we use data from simulations of noisy quantum computers to investigate the circumstances under which CNNs can accurately predict $s(c)$. Markovian stochastic Pauli errors--such as uniform depolarization or dephasing---are ubiquitous in experimental quantum computing systems and their effects are relatively simple to model. So we first study whether CNNs can model $s(c)$ in the presence of Markovian stochastic Pauli errors. In this section we show that CNNs can learn to accurately predict the success probabilities of few-qubit random circuits ($n=5$) that are subject to local stochastic Pauli errors. We demonstrate that, given sufficient data, a CNN will outperform an ERM fit to the same data. We explore the impact of dataset size, and we find that (1) increasing the training dataset size improves the CNNs predictions, and (2) CNNs perform well even with fairly small training datasets (e.g., $\sim$1000 circuits).

\subsection{Error models and datasets}
We constructed a dataset consisting of 16600 randomized mirror circuits (see Section~\ref{ssec:circuit-generation}), for a 5-qubit device with a ``T'' topology:
\begin{equation*}
\begin{tikzpicture}[node distance={5mm}, thick, main/.style = {draw, circle}] 
\node[main] (1) {}; 
\node[main] (2) [right of=1] {}; 
\node[main] (3) [right of=2] {}; 
\node[main] (4) [below of=2] {}; 
\node[main] (5) [below of=4] {}; 
\draw[-] (1) -- (2); 
\draw[-] (2) -- (3); 
\draw[-] (2) -- (4); 
\draw[-] (4) -- (5); 
\end{tikzpicture} 
\end{equation*}
The circuits varied in width from $1$ to $5$, with a circuit of width $w$ designed for and applied to a randomly chosen set of $w$ connected qubits. The circuits varied in depth from 3 to 1825 layers \footnote{Note that ``depth'' here and throughout means the number of circuit layers in the circuit, rather than the ``benchmark depth'' of the randomized mirror circuits (as defined in Ref.~\cite{Proctor2021-wt}), which is an alternative notion of the depth of a randomized mirror circuit that is adopted when using these circuits to estimate average layer error rates.}. 

We simulated the circuits under a single error model, consisting of local stochastic Pauli errors that are maximally biased: for each gate on each qubit, only one of $X$, $Y$ or $Z$ errors occurs with non-zero probability. The exact error model was randomly selected, i.e., which error can occur for a particular gate and qubit, and the rate of that error, was chosen at random (see Appendix~\ref{app:noise-models} for the selection protocol). We chose to simulate \emph{biased} errors because larger bias makes the task of modelling $s(c)$ harder in the following sense: when Pauli stochastic errors are biased the success probability of a circuit $s(c)$ not only depends upon the number of times each gate appears in a circuit, but also on the state of the qubits when that gate is applied (e.g., a $Z$ error has no impact on a qubit in a $Z$ eigenstate). The one-qubit and two-qubit error rates were selected to ensure a wide distribution of success probabilities $s(c)$. Each sampled circuit $c$ was simulated under the selected error model to compute its exact success probability $s(c)$, resulting in the dataset $D=\{(c,s(c))\}$. This dataset was then randomly partitioned into training, validation, and testing subsets---with a split of 70\%, 20\%, 10\%, respectively.

To explore how model performance depends on the amount of training data, and on the number of times each circuit was run, we used $D$ to create datasets with fewer total circuits, and with estimates of each $s(c)$ calculated from a finite number of repetitions ($N_{\rm shots}$) of each circuit $c$ \footnote{The exact value of $s(c)$ can be calculated when simulating an error model, but in experiments $s(c)$ can only ever be estimated from a finite number of runs of a circuit---each of which either returns the ``success'' bit string or does not.}. We created 11 instances of circuit sets containing 100 circuits (i.e., 70 training, 20 validation, and 10 test circuits) and 5 instances of circuit sets containing 300, 500, and 1000 circuits, by sub-sampling from our 16600 circuits \footnote{For each of the 5 (alt. 11) instances the datasets are nested, e.g., the 500 circuits are sampled from the 1000 circuits, and we use the same training, validation, and testing partition.}. This results in datasets in which the number of combined training and validation circuits ($N_{\textrm{circuits}}$) is equal to 90, 270, 450, 900, and 14940. For each dataset size, we created datasets with $N_{\rm shots}=100$, 1000, 10000, as well as datasets without shot noise (denoted by $N_{\rm shots} = \infty$), i.e., datasets containing $s(c)$ rather than estimates for $s(c)$.

\begin{figure}
    \centering
    \includegraphics[width=8cm]{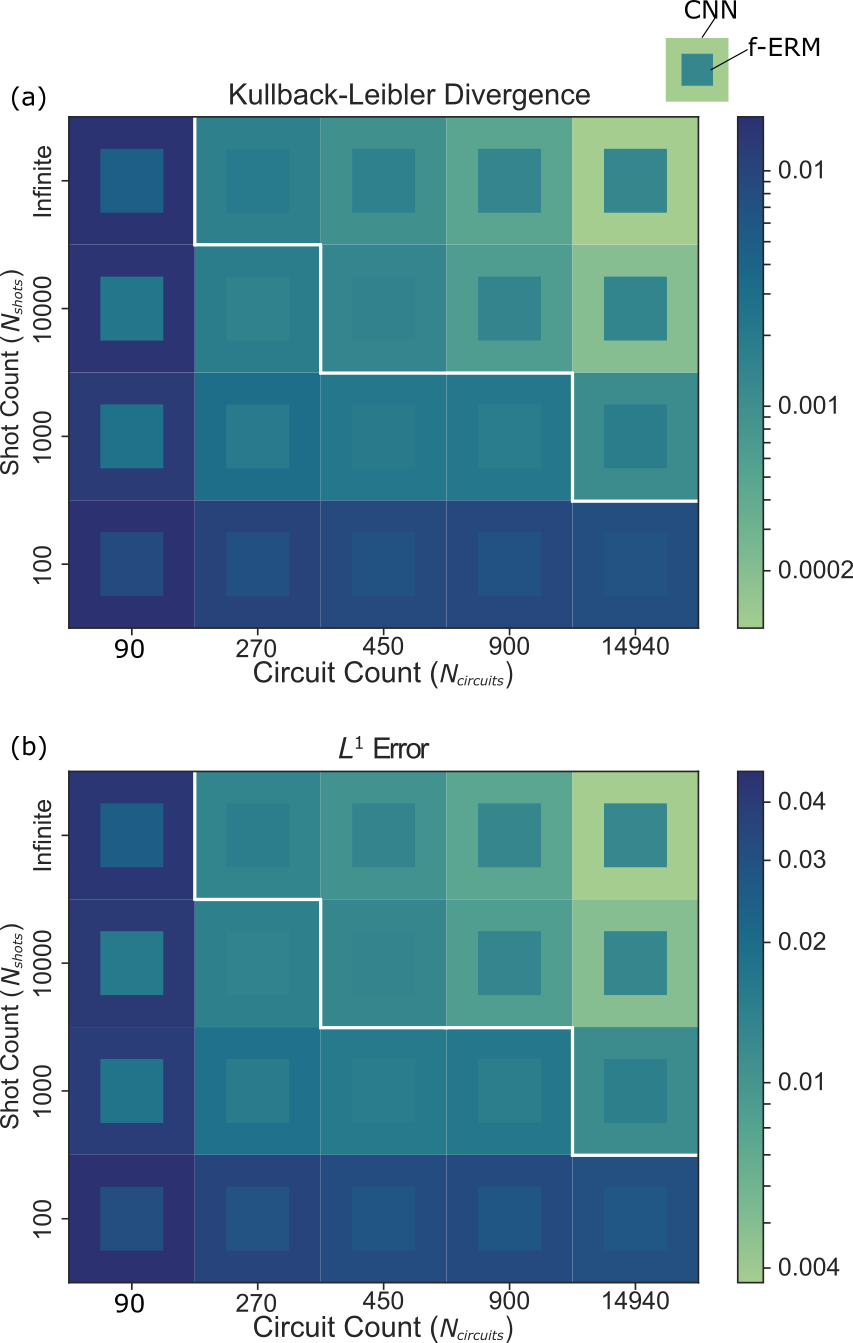}
    \caption{\textbf{Modelling the effect of stochastic errors with CNNs (cont.)} Quantifying the prediction error of CNNs and f-ERMs (fit error rates models) trained on simulated data from few-qubit random circuits with a stochastic Pauli errors model  (see the caption of Fig.~\ref{fig:markovian-predictions} for details). \textbf{(a)} The KL divergence and \textbf{(b)} the $L^1$ error for the CNN's predictions (outer squares) and f-ERM's predictions (inner squares), averaged over multiple randomly sub-sampled datasets of each size. The CNN's prediction accuracy surpasses that of the f-ERM in the region above the white line. In contrast with the f-ERM, we observe that the accuracy of the CNNs continues to increase up to the largest dataset size we used.}
    \label{fig:markovian-heatmaps}
\end{figure}

\subsection{Results}
For each dataset, we trained a CNN and fit an ERM on the same data. The prediction accuracy for the trained CNNs and the fit ERMs (f-ERMs) on the test data are summarized in Figs.~\ref{fig:markovian-predictions} and \ref{fig:markovian-heatmaps}. Each CNN's hyperparameters were tuned using the procedure described in Section~\ref{ssec:tuning}. For every dataset partition, we evaluated the performance of both the CNN and the f-ERM on the full set of test circuits (1660 circuits) using the true success probabilities $s(c)$ (i.e., the test data for $N_{\textrm{circuits}} = 14940$ and $N_{\rm shots} = \infty$). Fig.~\ref{fig:markovian-predictions} shows the predictions of trained CNNs and f-ERMs for a single instance of a dataset of each size and shot count. We observe that the prediction accuracy of the CNNs generally increases with both increased dataset size (increasing $N_{\textrm{circuits}}$) and reduced shot noise (increasing $N_{\rm shots}$). The CNN outperforms the f-ERM for sufficiently large $N_{\textrm{shots}}$ and $N_{\textrm{circuits}}$, although in our simulations the f-ERM outperforms the CNN a majority of the time.. An arguably necessary condition for a neural network model for $s(c)$ to be useful is that its prediction accuracy is better than a f-ERM (fitting ERMs is efficient and scalable, and an ERM's parameters can be interpreted), and we observe that our CNNs are satisfying this necessary criteria for sufficiently high-quality training datasets.

To quantify the accuracy of each model's predictions, in Figs.~\ref{fig:markovian-heatmaps} (a-b) we show the mean KL divergence ($d_{\textrm{KL}}$) and mean $L^1$ error ($d_{L^1}$) for each dataset size and shot count, averaged over the multiple dataset instances (at each value for $N_{\rm circuits}$ and $N_{\rm shots}$). The outer and inner squares show the prediction error for the CNN and f-ERM, respectively, as a function of $N_{\textrm{circuits}}$ and $N_{\textrm{shots}}$. The CNN's prediction error decreases with increasing training set size and shot count, as quantified by both the KL divergence and the $L^1$ error. For example, $d_{L^1} = 0.047$ and $d_{\rm KL} = 0.017$ averaged over the datasets with $N_{\textrm{circuits}} = 90$ and $N_{\textrm{shots}} = 100$, whereas $d_{L^1} = 0.08 $ and $d_{\rm KL} = 0.0006$ averaged over the datasets with $N_{\textrm{circuits}} = 900$ and $N_{\textrm{shots}} = 10000$. A moderately accurate ERM can be obtained by fitting to few data (see the blue data in the first column of Fig.~\ref{fig:markovian-predictions}). This is because (1) the ERM is a few-parameter model (in this case it has 28 parameters \footnote{These 28 parameters correspond to a readout error rate for each of the 5 qubits, $15=3 \times 5$ single-qubit gate error rates, and $8=4\times 2$ CNOT gate error rates (there are four edges in the connectivity graph, and CNOTs can be applied in each direction on the edge).}), and (2) the f-ERM captures significant aspects of the true, data-generating process (the f-ERM fails to capture the bias in the errors, but it does capture the average error in each gate when applied to a random input state). In particular, (when $n \gg 1$) the success rate $s(c)$ of a typical randomized mirror circuit $c$ under a stochastic Pauli error model is well-approximated (although not exactly modelled) by multiplying the probability of a gate causing no error over all the gates in a circuit. The f-ERM therefore significantly outperforms the CNN in the small dataset regime.

We find that the CNNs are more accurate than the ERMs when the datasets are moderately sized and have moderately low shot noise (see Fig.~\ref{fig:markovian-heatmaps} for the precise boundaries). An ERM cannot exactly represent the true data-generating process (a biased Pauli stochastic error model), so its performance is intrinsically limited even in the large dataset limit. These results imply that CNNs are able to learn features of circuits that are more predictive of circuit success probabilities than those encoded into an ERM (gate and readout error rates). The accuracy of the CNN models continues to increase up to the largest dataset size we used ($N_{\textrm{circuits}}$ in the combined training and validation sets). However, the prediction error will converge to a non-zero value as $N_{\rm circuits} \to \infty$ if the CNN's ansatz does not contain the exact $s(c)$ function (or if the optimizer cannot find this function).

\section{Predicting capabilities with many qubits and non-Markovian errors}\label{sec:49Q-sims}
Neural network approaches to modelling a quantum computer's capability $s(c)$ are appealing because it is plausible that they can circumvent some important limitations of conventional approaches to modelling $s(c)$. The conventional approach to predicting the success probability $s(c)$ of some circuit $c$ is to learn the parameters of a parameterized error model (e.g., process matrices with unknown entries) and to then predict $s(c)$ by simulating the circuit $c$ under this learned error model (e.g., by multiplying together the error model's process matrices). This approach has two important limitations: (1) a parameterized model evidently cannot account for effects that are outside of its model, and (2) it is often infeasible to compute $s(c)$ via simulation of $c$ under the learned error model, beyond the few-qubit regime. In this section, we explore whether neural network approaches to approximating $s(c)$ can avoid these two limitations. We investigate whether CNNs can accurately model $s(c)$ in (1) the many-qubit regime and (2) the presence of errors that cannot be described by the most common kinds of parameterized error models (i.e., error models that are restrictions on the maximal Markovian model, as defined in  Appendix~\ref{app:max-markovian-model}). Using simulated data, below we show that CNNs can learn to accurately predict the success probabilities $s(c)$ of many-qubit random circuits ($n=49$) that are subject to stochastic Pauli errors, and that accurate predictions are still possible with the addition of \emph{non-Markovian} stochastic errors.

\subsection{Datasets}\label{ssec:49Q-datasets}
One of the aims in this section is to explore whether CNNs can accurately predict $s(c)$ outside of the few-qubit setting. We therefore focus on a hypothetical 49-qubit system, with a $7 \times 7$ grid connectivity. We again consider the task of predicting $s(c)$ for randomized mirror circuits. We created a circuit set containing 10000 randomized mirror circuits, for our hypothetical 49-qubit device, with a training-validation-test split of 37.5\%, 12.5\%, and 50\% (we limited the number of training and validation circuits to 5000 to speed up the model training and hyperparameter tuning). The circuit widths ranged from 1 to 49 qubits, and a $w$-qubit circuit is designed for a randomly selected set of $w$ connected qubits. The circuit depths ranged from 4 to 272 layers.

\begin{figure}
\includegraphics[width=8cm]{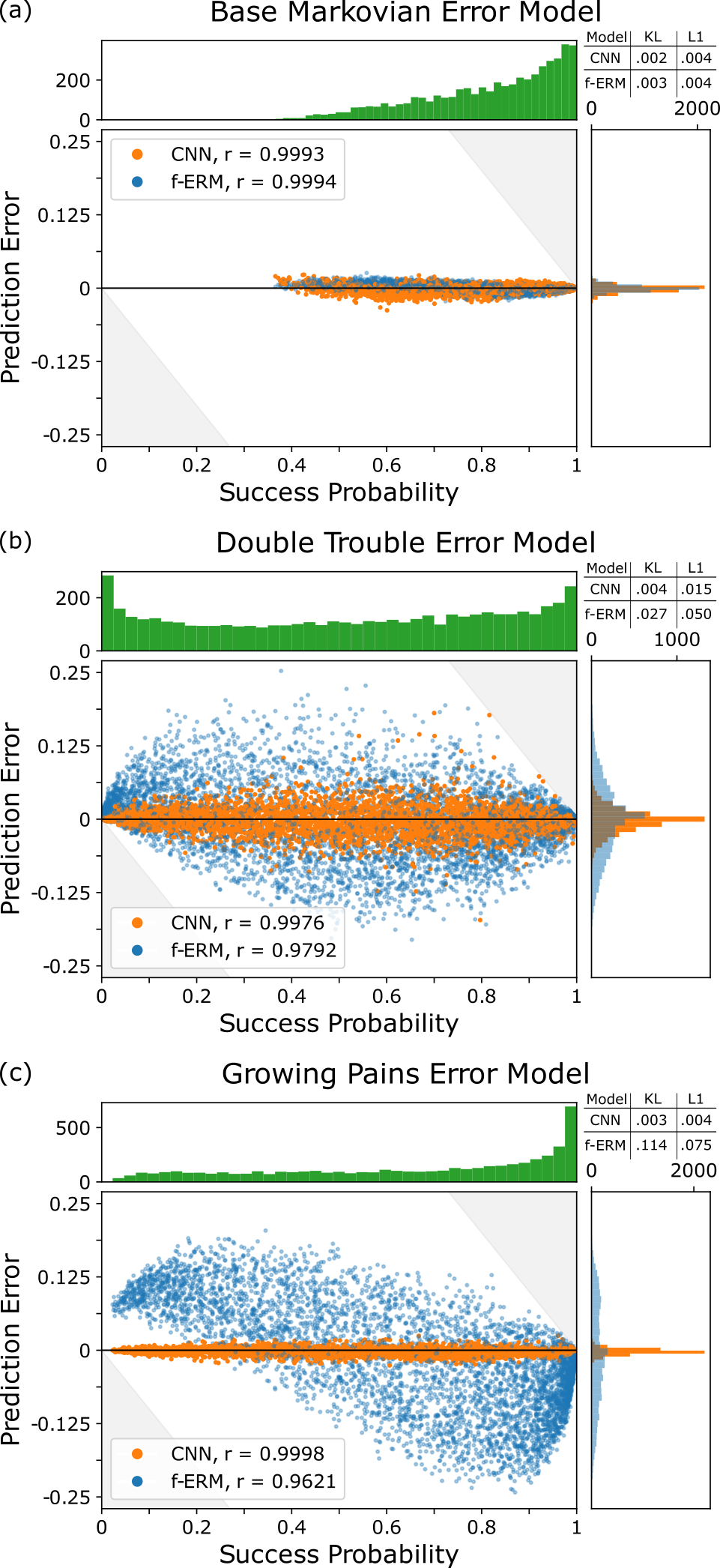} 
\caption{\textbf{Predictions for non-Markovian errors.} The prediction accuracy of CNNs trained on simulated data from random circuits for a hypothetical 49-qubit system with three different models. \textbf{(a)} A Pauli stochastic error model, with gate- and qubit-dependent error rates. This model forms the basis for two non-Markovian models in which \textbf{(b)} a two-qubit gate's error rate increases if it is preceded by a two-qubit gate on either of its qubits, and \textbf{(c)} gate errors increase over the duration of a circuit. Each main plot within (a-c) shows $s(c)$ versus the prediction error [$\delta(c)$ defined in Eq.~\eqref{eq:predict-error}] on test data, for a CNN and an ERM fit to the same dataset (f-ERM). The CNN and f-ERM have similar prediction error for the Markovian model, but the CNN significantly outperforms the f-ERM for both non-Markovian models. This is summarized by the $\delta(c)$ histograms [lower right, (a-c)] as well as the KL divergence and $L^1$ error for each model [upper right, (a-c)]. Each subfigure also contains a histogram of the test data [upper left, (a-c)].}
\label{fig:non-markovian-predictions}
\end{figure}

\subsection{Predicting many-qubit circuits}\label{ssec:base-markovian}
First we explore whether CNNs can predict circuit success probabilities in the many-qubit setting. To address this question we generated a many-qubit dataset by simulating the 49-qubit circuit set (described above) under a local stochastic Pauli error model with randomly selected error rates (the model parameters). We used this kind of error model for two reasons: (1) it enables direct comparisons with the performance of CNNs trained on the 5-qubit data presented in Fig.~\ref{fig:markovian-predictions}, and (2) it isolates the problem of predicting $s(c)$ for larger circuits $c$ from the problem of predicting more complex kinds of errors, e.g., non-Markovian stochastic errors (see Sections~\ref{ssec:growing-pains} and~\ref{ssec:double-trouble}) or coherent errors (see Section~\ref{ssec:coherent}). In this error model, each gate on each qubit is assigned independent error rates for the three possible Pauli errors, resulting in a data-generating model that is described by $1498$ parameters \footnote{There are $42$ edges in the graph, there is a CNOT gate associated with each direction on each edge, and each CNOT gate is associated with 6 error rates. There are $3\times 49$ different single qubit gates, each associated with 3 errors rates, and there are 49 readout error rates. This results in $42 \times 2 \times 6 + 4 \times 49 = 1498$ parameters in the model.} (the specific error model used is provided in the supplementary data and code \cite{dataset}). We can denote the parameters of this model by $\epsilon_P(g,i)$ where $P$ indexes the Pauli error ($X$, $Y$, or $Z$), $g$ denotes the gate (a single-qubit gate or CNOT, implicitly index by the qubit[s] on which it acts), and $i$ denotes one of the qubits that $g$ can act on. We computed $s(c)$ using a simulator that uses the approximation that two or more stochastic errors never cancel (see the supplemental data and code for details \cite{dataset}). This approximation enables fast simulation, and it is a very good approximation for random circuits with low gate error rates \cite{Polloreno2023-xa} as is the case here. We trained a CNN and fit an ERM using this dataset. For all datasets presented in this section, the CNN's hyperparameters were tuned using the procedure described in Section~\ref{ssec:tuning} (the hyperparameter space that we used is provided in the supplementary data and code \cite{dataset}).

Figure~\ref{fig:non-markovian-predictions} (a) shows the prediction error $\delta(c)$ on the test data for both the CNN and the f-ERM. The prediction error is simply 
\begin{equation}\label{eq:predict-error}
 \delta(c) = \hat{s}(c) - s_{\mathcal{M}}(c),
\end{equation}
where $s_{\mathcal{M}}(c)$ is the model's prediction.  The CNN's prediction error is small ($d_{L^1} = 4.34\times 10^{-3}$), demonstrating that CNNs can accurately predict success probabilities for many-qubit circuits and that the CNN can be trained using data from a practically feasible number of circuits (5000 circuits). This CNN's prediction error is similar to that observed when we trained a CNN on simulated data from a hypothetical 5-qubit system with a similar error model (see Fig.~\ref{fig:markovian-predictions}), while the number of parameters required to describe the true data-generating process has increased to around 1498. However, note that the f-ERM exhibits nearly the same prediction error as the CNN ($d_{L^1} = 4.18\times 10^{-3}$), demonstrating that this prediction problem is relatively easy. Furthermore, note that these results demonstrate that a CNN can be successfully trained on many-qubit circuits to predict $s(c)$ for one category of errors---local stochastic Pauli errors---but they do not imply that CNNs will be able to accurately model the effects of more complex kinds of errors that plague many-qubit quantum computing systems, such as crosstalk. 

\subsection{Non-Markovianity: temporal context dependence}\label{ssec:growing-pains}
We now investigate whether CNNs can accurately predict circuit success probabilities in the presence of effects that cannot be modelled by conventional parameterized models for quantum computers (including ERMs). Conventional parameterized models are constructed by placing restrictions on the maximal Markovian model (see Appendix~\ref{app:max-markovian-model}), so they cannot model non-Markovian errors. We conjecture that neural network methods for modelling $s(c)$ can learn good approximations to $s(c)$ even in the presence of many kinds of non-Markovian errors. To test this conjecture, we created a dataset for each of two different non-Markovian models. Each of these models is a modification to the Markovian error model described above, in Section~\ref{ssec:base-markovian}, and inspired by the non-Markovianity typically revealed by gate set tomography experiments~\cite{moueddene2022contextaware}.

One kind of non-Markovianity is gates that get worse over the course of a circuit (e.g., this can occur in ion-trap systems due to heating). To investigate whether CNNs can accurately model $s(c)$ in the presence of this kind of error, we created a simple model (``Growing Pains") where the error rate of each gate monotonically increases as a function of layer depth. 
In particular, for each error rate $\epsilon_{P}(g,i)$ in the Markovian model (see above), we replace the static error rate with an error rate that is indexed by the layer in the circuit in which the gate occurs:
\begin{equation}
    \epsilon_{\textrm{GP},P}(g, i, l)=  \frac{2\epsilon_P(g, i) + \epsilon^{\max}_P(g, i)[e^{l\tau} - 1]}{e^{l\tau} + 1},\label{eq:growing-pains}
\end{equation}
where $l$ denotes the layer index. Here $\epsilon_P(g, i)$ is the rate of the Pauli error $P=X,Y,Z$ when the gate at circuit location $(i,j)$ is applied to qubit $i$ in the base Markovian error model [note that $\epsilon_{\textrm{GP},P}(g, i, 0)=\epsilon_{P}(g,i)$], $ \epsilon^{\max}_P(g, i)$ is a parameter that specifies the error rate at $l \to \infty$ [we choose $ \epsilon^{\max}_P(g, i)=9\epsilon_P(g, i)$], and $\tau$ is a parameter that controls the rate of increase in the error rates (we choose $\tau=\nicefrac{1}{350}$) \footnote{For our choice for the values of the parameters in Eq.~\eqref{eq:growing-pains}, $\epsilon_{\textrm{GP},P}(g, i, l)$ is approximately linear in $l$ over the range of depths in our circuit set ($d$ up 272), with $\epsilon_{\textrm{GP},P}(g, i, 272)\approx 4 \epsilon_{\textrm{GP},P}(g, i, 0)$.}. We created a dataset for the Growing Pains error model, by simulating the set of 10000 circuits described in Section~\ref{ssec:49Q-datasets} to compute each circuit's $s(c)$, and we tuned and trained a CNN on this data (and fit an ERM).

Figure~\ref{fig:non-markovian-predictions} (c) shows the prediction error for the CNN and f-ERM, on the test data for the Growing Pains error model. The CNN's prediction error is low ($d_{\textrm{KL}} = 2.65\times10^{-3}$ and $d_{L^1} = 4.22\times10^{-3}$), and it is similar to the CNN's prediction error on the Markovian model ($d_{\textrm{KL}} = 1.82\times 10^{-3}$ and $d_{L^1} = 4.34\times 10^{-3}$). We find that $95\%$ of the CNN's predictions fall within $\pm 0.012$ of $s(c)$, with consistently good performance across circuits of different depths, widths, and values for $s(c)$. Convolutional layers are translationally invariant---each convolutional filter does not distinguish between the same gate pattern that appears near the start of circuit and near the end of a circuit---but the data generating model is not. However, some circuit location information is preserved by the convolutional layers of the network (the convolutional portion of the network outputs an image). Therefore, the subsequent dense network can learn weights that enable the complete network to model the effect of increasing gate error rates.

The CNN vastly outperforms the f-ERM---the $L^1$ error for the f-ERM ($d_{L^1}=0.075$) is almost twenty times larger than the $L^1$ error for the CNN trained on the same data. No conventional Markovian error model, including an ERM, can describe gates whose performance gets worse over the duration of a circuit. So, when fit to this dataset (which contains circuits of various depths, making this non-Markovianity visible), the best-fit error rates produce a f-ERM with over-optimistic predictions for deep circuits and over-pessimistic predictions for shallow circuits. This is the cause of the bias in the predictions of the f-ERM for high and low success probabilities circuits seen in Fig. \ref{fig:non-markovian-predictions} (b) [$s(c)$ is anti-correlated with circuit depth].

\subsection{Non-Markovianity: serial context dependence}\label{ssec:double-trouble}
We now apply CNNs to learn a capability function in the presence of another kind of non-Markovian error: \emph{serial context dependence}. In an error model with serial context dependence, a gate's error process depends on the gates that precede or follow it. To investigate whether CNNs can model $s(c)$ in the presence of serial context dependence, we constructed a simple model (``Double Trouble'') for this kind of error. We modified our Markovian error model so that a qubit's error rates for a two-qubit gate are increased if that qubit is acted on by another two-qubit gate in the preceding layer. Specifically, a gate $g$'s rate of Pauli $P$ errors on qubit $i$ in layer $l$ of circuit $c$ is given by
\begin{equation}
    \epsilon_{\textrm{DT},P}(g, i ,l) =  1 - [1 - \epsilon_P(g, i)] (1 - \epsilon_{\textrm{add}})^{N_{\textsc{cnot}}(i, l)N_{\textsc{cnot}}(i, l-1)},\label{eq:double-trouble}
\end{equation}
where $N_{\textsc{cnot}}(i, l) = 1$ if location $(i,l)$ in circuit $c$ is a CNOT gate and otherwise $N_{\textsc{cnot}}(i, l) = 0$. We choose $\epsilon_{\textrm{add}} = 0.005$, so if a qubit is involved in two consecutive CNOT gates, the error rate of the second CNOT gate increases by approximately 0.005.

Figure~\ref{fig:non-markovian-predictions} (b) shows the prediction error for a CNN, trained and tuned on data from the Double Trouble error model, and a ERM fit to the same data (f-ERM). The CNN's prediction error on test data is significantly larger than it is for a CNN trained and tested on the base Markovian model (the $L_1$ error is approximately three times larger: $d_{L^1} = 0.015$ compared to $d_{L^1} = 4.34\times 10^{-3}$). However, the CNN still significantly outperforms the f-ERM, as the CNN's $L_1$ error is approximately three times smaller ($d_{L^1} = 0.015$ compared to $d_{L^1} = .05$). We find that $95\%$ percent of the CNN's predictions are within $\pm 0.045$ of $s(c)$. 

Convolutional layers can pick out localized pattern of gates, by learning convolutional filters (and biases) that return a non-zero value if and only if that pattern of gates appears. Sequential CNOT gates can therefore be identified by a convolutional filter applied directly to the input image representation of a circuit (i.e., a convolutional filter in the first convolutional layer). In particular, there exists a set of four convolutional filters that enable a convolutional layer to identify all instances of sequential CNOT gates (these filters are provided in Appendix~\ref{app:filters}). This suggests that the CNN training and hyperparameter tuning process that we have used is finding significantly sub-optimal convolutional filters in this case.

\section{Challenges to useful neural network models for capabilities}\label{sec:challenges}
In this section we explore some important challenges to creating useful neural network models for a quantum computer's capability $s(c)$. We explore the problem of predicting $s(c)$ for circuits sampled from a different distribution to the training data (Sections~\ref{ssec:ood-wide-predictions}); we highlight the difficulty of modelling the impact of coherent errors using neural networks (Section~\ref{ssec:coherent}); and we investigate the importance of including error sensitivity information within the neural network's input (Section~\ref{ssec:no-stabilizers}).

\subsection{Generalizing to out-of-distribution circuits}\label{ssec:ood-wide-predictions}
Each practical application for a model of $s(c)$ will require that the model's predictions are accurate for some set of circuits of interest $\mathbb{C}$, and this circuit set will generally be application-specific. For example, one application for a model of $s(c)$ is finding a low-error compilation for some algorithm $\mathcal{A}$. There are many circuits that implement an algorithm $\mathcal{A}$, and the primary aim in compilation is to find the circuit $c$ in the set of all such circuits ($\mathbb{C}_{\mathcal{A}}$) that maximizes $s(c)$. Using a model for $s(c)$ to inform this compilation (e.g., to define a cost function to be used by an optimizer) requires that it is accurate for the circuit set $\mathbb{C}_{\mathcal{A}}$. The relevant set of circuits will vary between applications of our model for $s(c)$. For each circuit set $\mathbb{C}$ of interest, a neural network can be trained on data from circuits sampled from $\mathbb{C}$. However, retraining a neural network is expensive (it requires new training data, and network training can be slow). Therefore, it is interesting to explore whether neural network models can accurately predict $s(c)$ for circuit sets $\mathbb{C}$ that are sampled from a different distribution to the training data---a task that is often referred to as \emph{out-of-distribution generalization} \cite{She21}. Below we present two examples of out-of-distribution generalization.

\begin{figure}[t!]
\includegraphics[width=8cm]{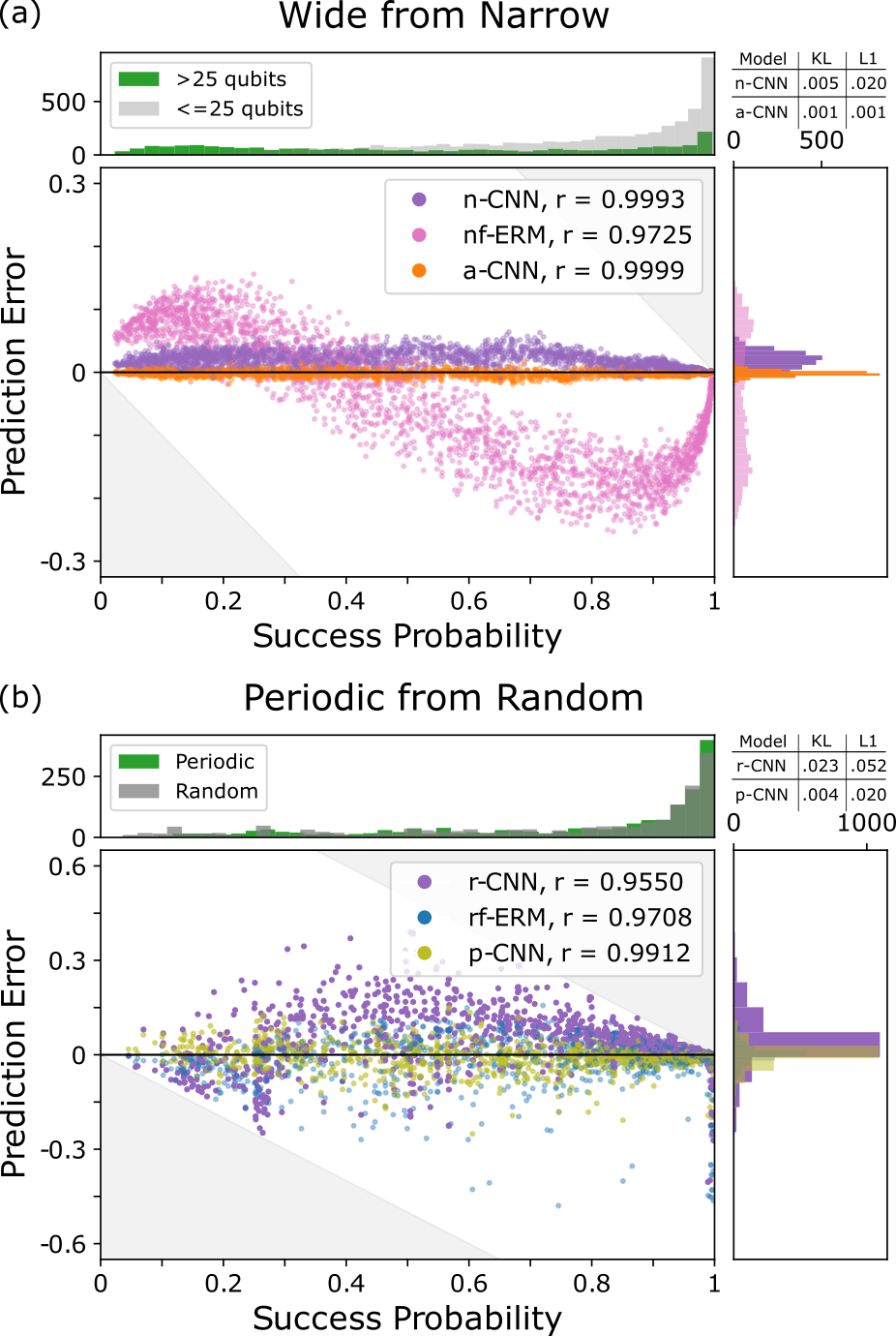} 
\caption{\textbf{Generalizing to out-of-distribution circuits.} Two examples of generalizing a neural network to predict circuits that are drawn from a distribution that differs significantly from the training distribution. \textbf{(a)} The prediction accuracy [$\delta(c)$] of a CNN (n-CNN, purple) trained on narrow circuits ($w \leq 25$ qubits) evaluated on a test dataset containing wide circuits ($w > 25$ qubits). The data was simulated under an error model in which gate errors increase with circuit depth, so narrow circuits reveal all important aspects of the error model, meaning that accurate generalization to wider circuits is feasible. The prediction error $\delta(c)$ for n-CNN is much smaller than for an ERM fit to the same training data (nf-ERM, pink). However, $\delta(c)$ is larger for n-CNN than for a CNN trained on a dataset containing both narrow and wide circuits (a-CNN, orange). \textbf{(b)} The prediction error of a CNN (r-CNN, purple) trained on random circuits and evaluated on periodic circuits. The r-CNN has an $L^1$ error of $d_{L^1} = 0.005$ on test data drawn from the same distribution as the training data (random circuits), but it increases to $d_{L^1} = 0.052$ on periodic circuit data, resulting in worse predictions than provided by an ERM fit to the same training data (rf-ERM, blue). However, the CNN's performance can be substantially improved by re-training the CNN on periodic circuit training data (p-CNN, yellow), while maintaining the same architecture that was found by hyperparameter tuning using the random circuits data.}\label{fig:scatter-histogram-out-of-distribution}
\end{figure}

First, we consider the problem of predicting wide circuits (here $w >25$ active qubits) using a CNN trained on data containing only narrow circuits ($w \leq 25$ active qubits). This is a simple example with which to explore out-of-distribution generalization, but it also has practically relevance. In particular, for some definitions for $s(c)$ (such as TVD) there is no known method for efficiently measuring $s(c)$ in an experiment if $c$ is a wide and deep circuit (see Section~\ref{sec:capability}). We use the data simulated under the Growing Pains noise model (see Section~\ref{sec:49Q-sims}). We trained a CNN on only the narrow circuits in the training dataset (n-CNN), and we fit an ERM to the same data (nf-ERM). No automated hyperparameter tuning was performed, except to select the number of training epochs (see the supplemental data and code \cite{dataset} for the architecture) \footnote{A validation set of narrow circuits was used.}. In this error model, gate performance gets worse later in a circuit, but a gate's error rate is independent of width. In particular, the parameters of the data generating error model (Growing Pains) can be learned from the $w \leq 25$ qubit circuits data. It is therefore plausible that a CNN trained on data from narrow circuits will generalize to accurately predict $s(c)$ for wide circuit (whereas we could not expect a model trained on few-qubit circuits data to accurately predict the success probabilities of many-qubit circuits if there are additional errors that only appear in those many-qubit circuits, e.g., many-qubit crosstalk effects \cite{Hines2022-xv}). 

Figure \ref{fig:scatter-histogram-out-of-distribution} displays the n-CNN's out-of-distribution predictions on the wide circuit test data (purple points) alongside the predictions of the CNN that was trained on all of the training data (a-CNN, orange points), which includes narrow and wide circuits. The n-CNN generalizes moderately well in the sense that the prediction error of n-CNN is reasonably small ($d_{L^1}=0.02$). However, the prediction error for the n-CNN is approximately twenty times larger than for a-CNN ($d_{L^1}=0.001$), so the relative increase in the prediction error when removing wide circuits from the training dataset is large. There is also a bias in the out-of-distribution network's predictions: the n-CNN systematically underestimates $s(c)$ on the out-of-distribution circuits [i.e., typically $\delta(c) > 0$]. Interestingly, the n-CNN still achieves low prediction error on those test circuits with small $s(c)$, even though circuits with small $s(c)$ are underrepresented in the narrow circuit training data. 

Our second example of out-of-distribution generalization considers the problem of predicting $s(c)$ for circuits with different structures to those in the training dataset. In this work we have so far considered training CNNs on data from \emph{random} circuits. In particular, we have used data from randomized mirror circuits (see Section~\ref{sec:capability}), and these circuits are defined by a distribution that has support on all possible mirror circuits. However, a circuit sampled from this distribution is almost certainly highly disordered \cite{Proctor2021-wt}, e.g., it will almost certainly not contain repeated patterns of gates. It is not \emph{a priori} clear whether a CNN trained only on highly disordered circuits (e.g., randomized mirror circuits) will generalize to accurately predict $s(c)$ for highly ordered circuits (e.g., periodic circuits, or algorithmic circuits). However, error propagation in disordered circuits (where errors are scrambled, and coherent errors add quadratically) \cite{Polloreno2023-xa} differs substantially from error propagation in highly ordered circuits (where errors can be amplified or echoed away, and coherent errors can add linearly) \cite{Nie21}, suggesting that neural networks trained only on disordered circuits will generalize poorly to structured circuits.

To explore whether CNNs trained on random circuits generalize to ordered circuits we created a dataset of 5-qubit periodic mirror circuits and simulated them under the Markovian error model described and used throughout Section~\ref{sec:5Q-sims} (our simulation used $N_{\text{shots}} = \infty$ shots). We used a CNN that was trained on data from $N_{\text{circuits}} = 14940$ random mirror circuits (also with $N_{\text{shots}} = \infty$) to predict $s(c)$ for these periodic mirror circuits (we denote this CNN by r-CNN). Figure~\ref{fig:scatter-histogram-out-of-distribution} (b) shows r-CNN's prediction error on the test dataset (i.e., the periodic mirror circuits). The prediction error for r-CNN is large and it is approximately ten times larger on this out-of-distribution test data ($d_{L^1}=0.052$) than for in-distribution test data ($d_{L^1} \approx 0.004$).

To quantify the performance of the r-CNN we compare it to two alternative models: an ERM fit to the random circuit data (rf-ERM) and a CNN trained on periodic mirror circuit data (p-CNN). As shown in Fig.~\ref{fig:scatter-histogram-out-of-distribution} (b), the rf-ERM outperforms the r-CNN on the out-of-distribution test data, even though CNNs trained on random circuit data have substantially lower prediction error than fit ERMs on in-distribution test data (see Fig.~\ref{fig:markovian-heatmaps}). However, if we retrain the CNN using the periodic mirror circuit training data, we obtain significantly better model performance (p-CNN in Fig.~\ref{fig:scatter-histogram-out-of-distribution}). These results suggest that CNNs learn features that encode how a specific error model interacts with the specific class of circuit used in the training. This is both a strength---as it enables CNNs to outperform ERMs, which cannot model the interaction between circuit structure and a specific error model---and a weakness---as it limits the prediction accuracy of CNNs on out-of-distribution circuits. We conjecture that two complementary approaches will improve a CNNs out-of-distribution prediction accuracy: (1) fine-tuning a trained neural network using a small amount of data for each circuit family of interest, e.g., by retraining only some of a network's weights; (2) encoding known physics for how errors propagate through circuits within a neural network's architecture and/or within the circuit encoding.

\begin{figure}[t!]
\includegraphics[width=8cm]{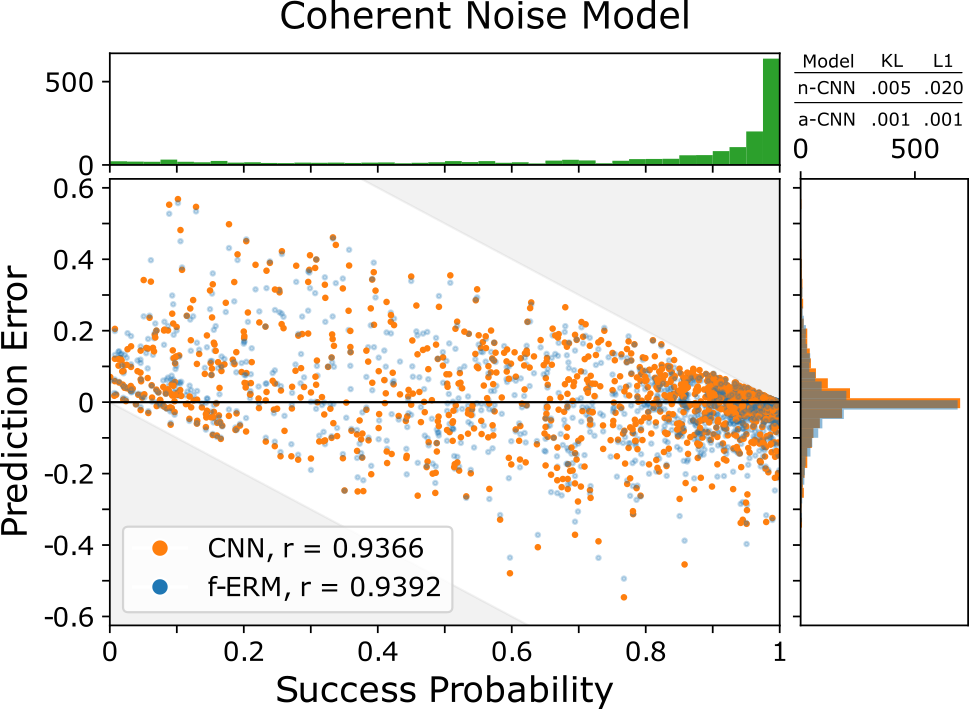} 
\caption{\textbf{Inaccurate capability models when errors are purely coherent.} The prediction accuracy of a CNN and a f-ERM trained on randomized mirror circuit data ($n=5$ qubits) generated from an error model with purely coherent (i.e., Hamiltonian) errors. We find that neither model accurately predicts the test data ($d_{L^1} \approx 0.06$ for both models). Predicting $s(c)$ in the presence of coherent errors is challenging because coherent errors can add or cancel across an entire circuit.}
\label{fig:coherent-errors}
\end{figure}

\subsection{Prediction in the presence of coherent errors}\label{ssec:coherent}
Techniques for modelling capability functions will only be useful in practice if they can learn a good approximation to $s(c)$ in the presence of all the types of errors that real processors commonly experience. In this paper so far, we have only considered modelling $s(c)$ in the presence of (Markovian and/or non-Markovian) stochastic Pauli errors, but real quantum processors also experience many other kinds of error (e.g., see Ref.~\cite{Mavadia2018-ki}). Coherent errors are ubiquitous and their effect on $s(c)$ is particularly challenging to predict, because they can coherently add or cancel within circuits. We therefore investigated whether CNNs can accurately model $s(c)$ in the presence of coherent errors.

We simulated a set of 5-qubit randomized mirror circuits (the circuits of Section~\ref{sec:5Q-sims}) under an error model consisting of purely coherent errors on each gate. We randomly sampled the error model, using the procedure provided in Appendix~\ref{app:noise-models}. Figure~\ref{fig:coherent-errors} shows the prediction accuracy for a CNN (with tuned hyperparameters) and a f-ERM. The CNN has a slightly larger prediction error ($d_{L_1} = .06$) than the f-ERM ($d_{L_1} = .0599$), and neither model's predictions are accurate. For both models, there are circuits within the test set for which the $L_1$ error is over 0.5, which is arguably a catastrophic prediction failure. Coherent errors are a significant proportion of the total error in many contemporary quantum computing systems (see, e.g., Refs.~\cite{Mavadia2018-ki,Hashim2022-pe}), and so CNNs' inability to model $s(c)$ in the presence of coherent errors will limit the prediction accuracy of CNN models for $s(c)$ that are trained on experimental data (this is consistent with our results in Section~\ref{sec:experimental-data}).

Low prediction accuracy for CNN models of $s(c)$ in the presence of coherent errors can be explained as follows. The impact of coherent errors on a particular circuit $c$ is difficult to predict because coherent errors can coherently add or cancel across the entire circuit. Whether two coherent errors at two circuit locations $(i,j)$ and $(i',j')$ coherently add or cancel strongly depends both on these errors (their magnitudes and directions) as well as the unitary evolution caused by the circuit layers between $j$ and $j'$. Furthermore, because coherent errors at any two circuit locations can cancel (or add), their combined effect likely cannot be modelled by a CNN that has convolutional filters that are much smaller than the circuit size. Exact classical modelling of the effect of coherent errors on a general circuit $c$ likely requires simulating the unitary evolution of each circuit layer, and it is perhaps infeasible for a neural network to learn a good approximation to $s(c)$ when coherent errors dominate. However, it is possible that improvements to the circuit encoding (see below) or the neural network architecture may greatly improve the accuracy of neural network models for $s(c)$ in the presence of coherent errors, as we discuss briefly in Sections~\ref{ssec:no-stabilizers} and~\ref{sec:discussion}. Furthermore, note that coherent errors can be converted into stochastic Pauli errors using randomized compiling \cite{Hashim2022-pe, Wallman2016-rd} or Pauli frame randomization \cite{Knill2005-xm}.

\subsection{The role of error sensitivities in capability learning}\label{ssec:no-stabilizers}
Our encoding [$I(c)$] for a circuit $c$ includes a limited amount of information about the sensitivity of that circuit to errors. In particular, pixel $I_{ij}(c)$ encodes information about whether a Pauli $X$, $Y$, or $Z$ error on qubit $i$ after layer $j$ will change the state of the qubits (see Section~\ref{ssec:encoding}). This information is stored in three ``error sensitivity'' channels, that correspond to these three kinds of error. We included these three channels in our encoding as we conjectured that they make it easier for a CNN to learn an accurate model for $s(c)$ in the presence of local stochastic Pauli errors (including non-Markovian local stochastic Pauli errors). This conjecture is supported by the observation that there exists a CNN (an architecture with specific weights) that is a good approximation to any local stochastic Pauli error model and that our construction for such a CNN (see Appendix~\ref{app:lps-filters}) uses the information in these error sensitivity channels. 

\begin{figure}[t!]
\includegraphics[width=8cm]{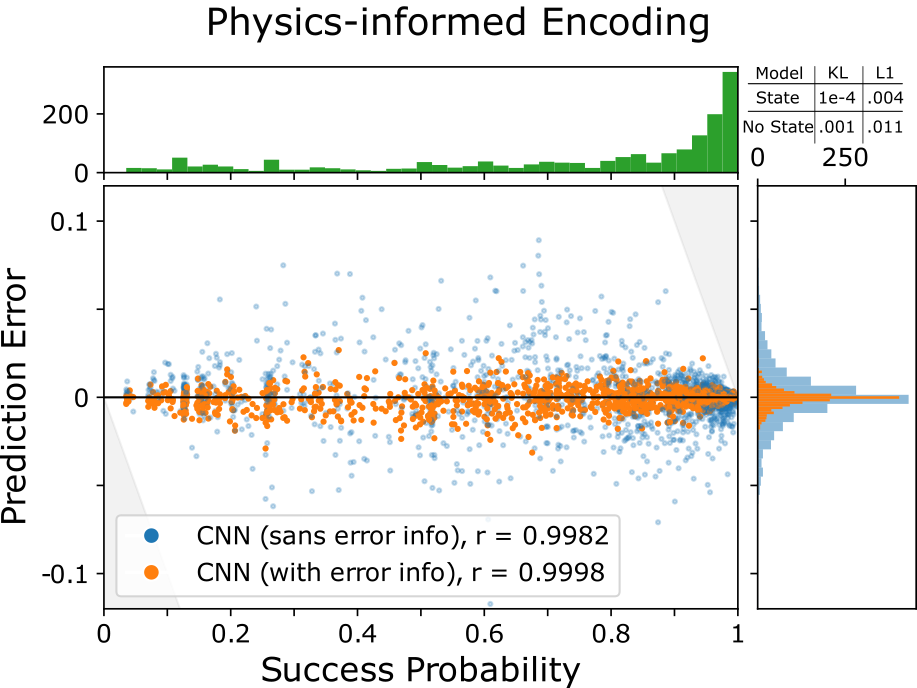}
\caption{\textbf{Error sensitivity information improves model accuracy.} The prediction error of CNNs trained with and without the error sensitivity channels---which we use to encode information about each qubit's sensitivity to the three single-qubit Pauli errors at each circuit location---on randomized mirror circuit data ($n=5$ qubits) simulated under a Pauli stochastic error model. We observe significantly better performance (the KL divergence is an order of magnitude smaller) for the CNN that has access to the error sensitivity channels. This suggests that error sensitivity information is important for accurate capability learning.}
\label{fig:scatter-stabilizer-vs-no-stabilizer}
\end{figure}

To examine the importance of the error sensitivity channel, we trained a CNN on a dataset that removed the error sensitivity channels from our circuit encoding $I(c)$. We used the 5-qubit local Pauli stochastic errors dataset from Section~\ref{sec:5Q-sims} (we used the $N_{\textrm{shots}} = \infty$ and $N_{\textrm{circuits}} = 14940$ dataset). We did not perform hyperparameter tuning (using instead the network architecture found when performing hyperparameter tuning with the error sensitivity channels included). Figure~\ref{fig:scatter-stabilizer-vs-no-stabilizer} shows the predictions of the CNN trained on datasets that do and do not include the error sensitivity channels. The performance of the CNN without access to the error sensitivity information is significantly worse ($d_{L^1} =0.011$ versus $d_{L^1} = 0.004$), supporting our hypothesis that this error sensitivity information significantly reduces the difficulty of the learning problem.

Our results suggest that channels that encode a circuit's sensitivity to local Pauli stochastic errors improve a CNN's ability to model $s(c)$ in the presence of local Pauli stochastic errors. This suggests a promising path forward for accurate modelling of $s(c)$ in the presence of more general classes of error, including coherent errors: the inclusions of additional error sensitivity information in the circuit encoding. There are, however, significant challenges to this approach. First, the impact of some kinds of errors---including coherent errors---cannot be localised to each circuit location, because their overall effect depends on how they combine. Second, our current method for including error sensitivity information within $I(c)$ relies on the circuit containing only Clifford gates---but a useful model for $s(c)$ arguably also needs to predict $s(c)$ for non-Clifford circuits. We therefore suggest that an interesting open problem is the development of a representation of circuits that includes partial or approximate error sensitivity information for a broad range of errors errors in general circuits.

\begin{figure*}[t!]
\includegraphics[width=17.2cm]{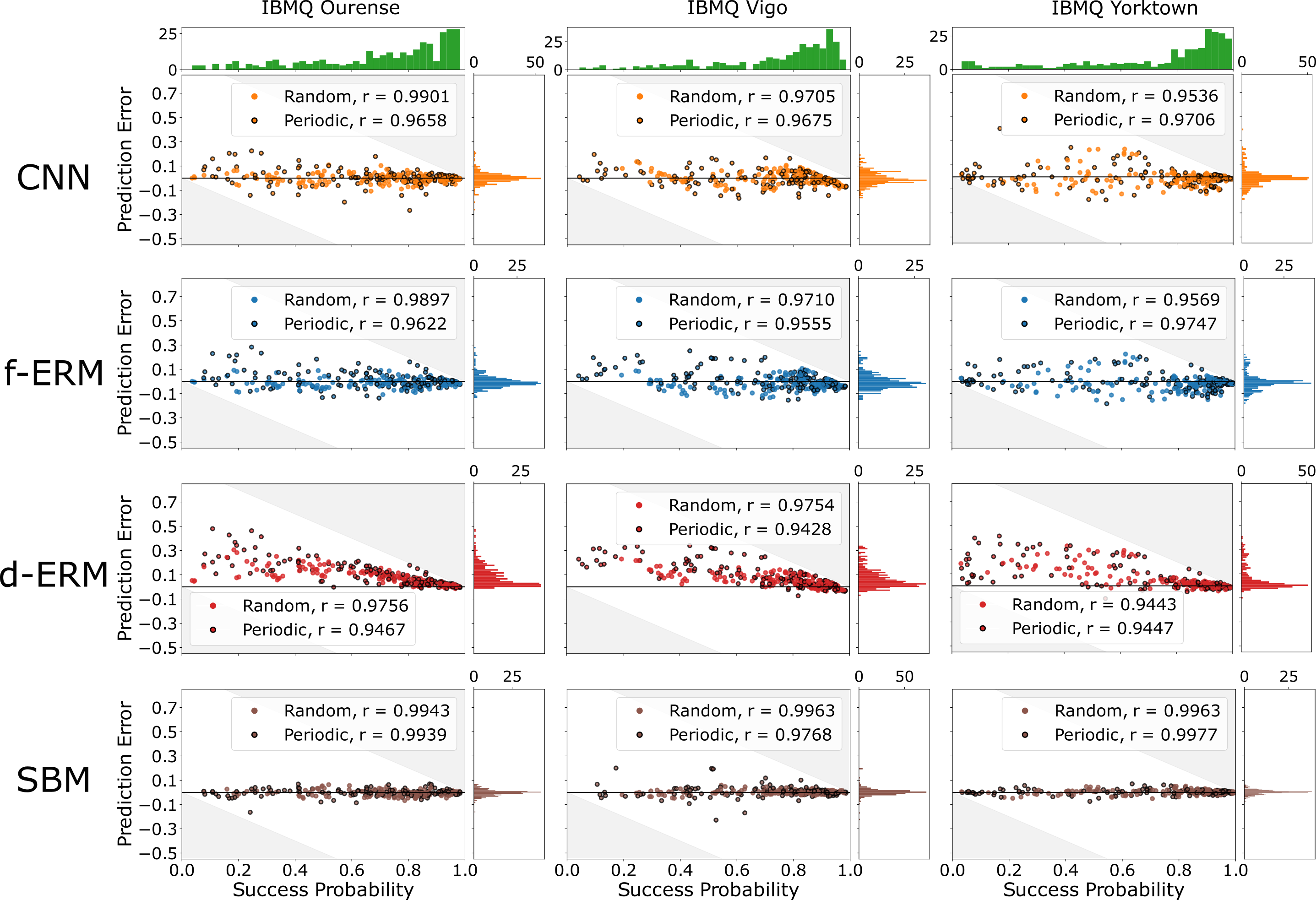} 
\caption{\textbf{Predicting the capabilities of cloud-access IBM Q processors.} We trained CNNs to predict circuit success probabilities for seven cloud-access IBM Q processors (which are all 5-qubit or 14-qubit systems), using a dataset that consists of success probability estimates $\hat{s}(c)$ for approximately 3000 (5-qubit processors) or 5000 (14-qubit processors) periodic and randomized mirror circuits of varying widths and depths. We show the prediction error on the test data (for three processors) for the CNN (top row, orange), an ERM fit to the same training data (f-ERM, second row, blue), an ERM that uses the gate error rates provided by IBM (d-ERM, third row, red), and a ``stability baseline model'' (SBM) that quantifies the stability of the processor (fourth row, brown). The CNN and f-ERM's prediction error are comparable and constitute moderate prediction accuracy, although the CNN outperforms the f-ERM in most cases (see Fig.~\ref{fig:experiment-metrics}). Prediction error is separated into periodic mirror circuits (black outline) and randomized mirror circuits (no outline), and for all but one processor we observe lower prediction error for the randomized mirror circuits (see $r$ values reported in legends). Both the CNN and f-ERM are substantially more accurate than the d-ERM (the d-ERM's error rates are known to miss important sources of error \cite{Hines2022-xv}). The SBM model (fourth row) is a baseline that quantifies how stable each circuit's success probability is over time (see main text for details), and it is unlikely that a CNN (or another model) will outperform the SBM without including additional context information in the training data (e.g., timestamps). We observe that neither the CNN nor the ERMs achieve the prediction accuracy of the SBM, but for all but two processors the average $L^1$ error of the CNN is less that two times the average $L^1$ error of the SBM (see Fig.~\ref{fig:experiment-metrics}).}
\label{fig:experimental-combined}
\end{figure*}

\begin{figure}[ht!]
\centering
\includegraphics[width=8cm]{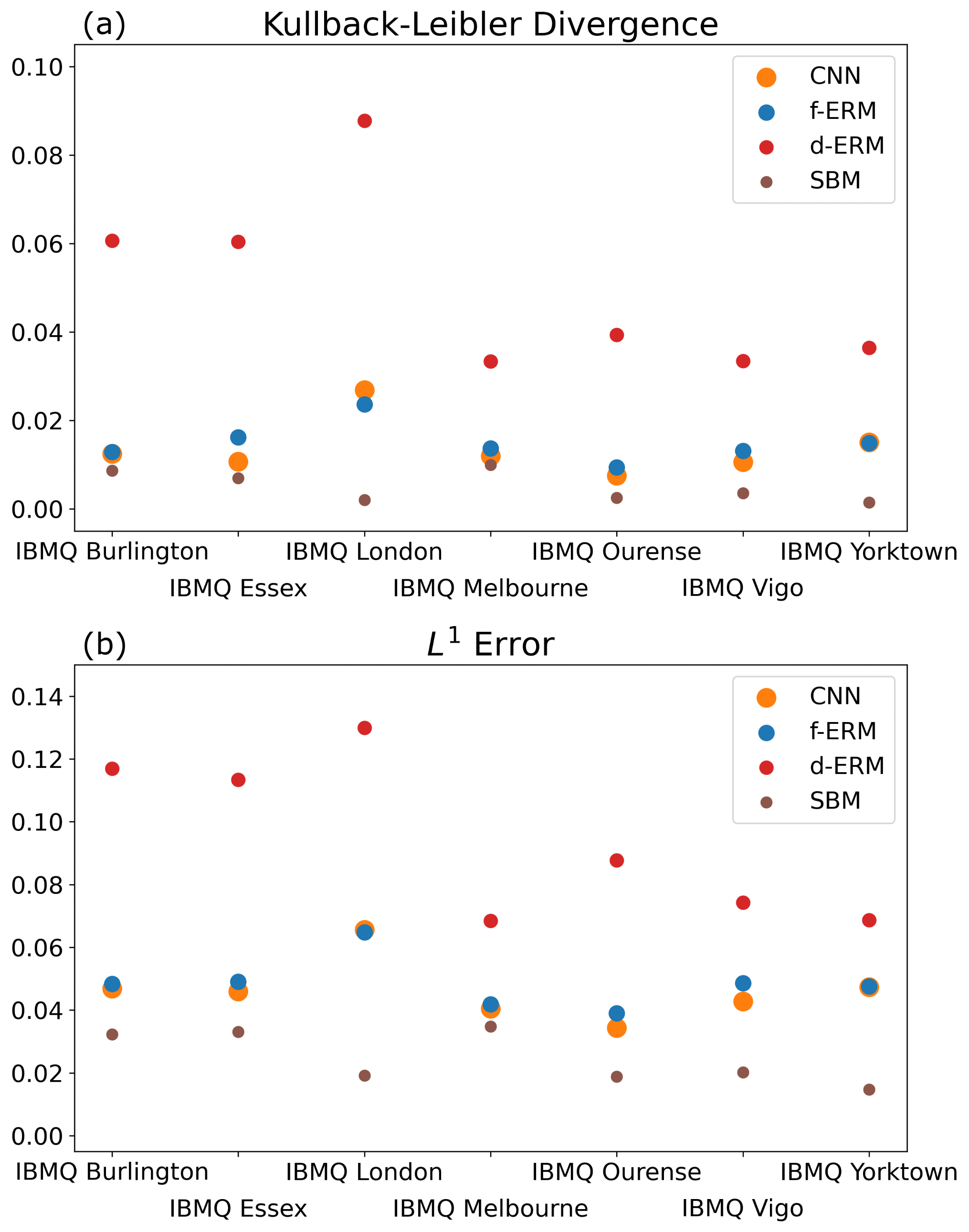} 
\caption{\textbf{Prediction accuracy of capability models for cloud-accessible IBM Q processors.}  The prediction accuracy on test data for CNNs trained on data from seven different cloud-access IBM Q processors, quantified by \textbf{(a)} KL divergence and \textbf{(b)} $L^1$ error. We also show the prediction accuracy for f-ERMs, d-ERMs, and SBMs. See Fig.~\ref{fig:experimental-combined} and Section~\ref{sec:experimental-data} for further details. (In all cases, error bars are too small to be visible.)}
\label{fig:experiment-metrics}
\end{figure}

\section{Demonstration on experimental data}\label{sec:experimental-data}
In this section we explore how accurately CNNs can model the capabilities of cloud-access quantum computing systems.

\subsection{Datasets}\label{ssec:experimental-datasets}
We used data from an existing publicly-available dataset \cite{mcqc-dataset} that consists of success probabilities for varied width and depth randomized and periodic mirror circuits, run on various cloud-access quantum processors. We used data from seven different IBM Q processors: a 14-qubit system (IBM Q Melbourne), and six different 5-qubit systems. Each processor's dataset was obtained by running 40 shape $(w,d)$ randomized mirror circuits and 40 shape $(w,d)$ periodic mirror circuits, with $w$ and $d$ varied systematically ($w$ took every possible value, and $d$ was  exponential spaced). For each processor, all $w$-qubit circuits were run on the same set of $w$ qubits. The number of circuits in each dataset is approximately $3000$ (for the 5-qubit systems) or approximately 5000 (for IBM Q Melbourne). Each set of circuits was run, in a randomized order, obtaining an estimate $\hat{s}(c)$ for each circuit's success probability $s(c)$ from  $N_{\rm shot} = 1024$ executions of the circuit. The entire circuit list was then immediately run again, obtaining a second estimate $\hat{s}'(c)$ (with $N_{\rm shot} = 1024$) for each circuit. We use the data from the first pass through the circuits (``pass 1'') for training and evaluating our CNNs, and the data from the second pass through the circuits (``pass 2'') to quantify stability. Further details of these datasets and experiments are presented in Ref.~\cite{Proctor2021-wt} (the datasets used here are referred to as ``experiment 2'' in the supplementary information therein \footnote{Experiment 2 consisted of running the same benchmark on seven IBM Q processors and one Rigetti processor. Herein we do not present results for the Rigetti processor. This is because, at the time of those experiments, Rigetti did not provide up-to-date calibration data, so we cannot create an error rates model from that data.}). We randomly separated the circuits for each processor into training, validation, and test circuits, with an 80\%, 10\%, 10\% split.

\subsection{Results}\label{ssec:experimental-results}
We trained and tuned CNNs separately for each processor, using that processor's training and validation datasets. Figure~\ref{fig:experimental-combined} (top row) shows the prediction accuracy of the CNNs on the test data for three of the seven processors, separated into periodic and randomized mirror circuits. Equivalent plots for the other four processors can be found in Appendix~\ref{app:experimental-predictions} (see Figs.~\ref{fig:app-experimental-predictions-1} and \ref{fig:app-experimental-predictions-2}). We observe moderate prediction accuracy---the average $L^1$ error on the test data ranges from 0.04 to 0.08, and the KL divergence ranges from 0.01 to 0.09 (see Fig.~\ref{fig:experiment-metrics}). This prediction accuracy is similar to or better than that obtained in other recent work that applies neural networks to model $s(c)$ on cloud-access quantum processors \cite{Wan22, Ame22}, but note that those papers apply neural networks to slightly different tasks and different circuit families, so meaningful quantitative comparisons are not possible. For all but one processor, the prediction error for the randomized mirror circuits is smaller than the prediction error on the periodic mirror circuits (compare the $r$ values in the legends of Figs.~\ref{fig:experimental-combined}, \ref{fig:app-experimental-predictions-1} and \ref{fig:app-experimental-predictions-2}). The success probabilities of periodic circuits are typically harder to predict, due to the interactions between the (unknown) structure in a processor's errors and the circuit's structure. Note that we observed lower prediction accuracy on periodic circuits than on random circuits with simulated data (see Section~\ref{ssec:ood-wide-predictions}).

To explore how the accuracy of our CNNs compares to alternative models for $s(c)$, we compare their predictions to two classes of ERMs. One of these ERMs has parameters (i.e., error rates) obtained from the processor's performance data provided by IBM Q \footnote{We use the same method as described in Ref.~\cite{Proctor2021-wt} to turn IBM Q's RB error rates into values for the parameters of an error rates model.}, which we call the ``device error rates model'' (d-ERM). The CNN substantially outperforms the d-ERM for every dataset, as shown in Fig.~\ref{fig:experimental-combined} (compare the top and third rows) and Fig.~\ref{fig:experiment-metrics}. This large improvement in the prediction accuracy compared to a d-ERM highlights the limitations of using parameterized error models obtained from only one- and two-qubit RB experiments (IBM Q's performance data), as they miss many important sources of error in many-qubit circuits (e.g., crosstalk). As throughout this paper, we also compare each CNN's predictions to an ERM whose parameters are also fit to the training data (f-ERMs). For most processors, the CNN's prediction error is lower than the f-ERM, but it is only a slightly improvement (see Fig. \ref{fig:experiment-metrics}, and compare the top and second row of Fig.~\ref{fig:experimental-combined}). For one dataset (from IBM Q London), the f-ERM actually outperforms the CNN, even though a good approximation to the f-ERM exists within the CNN's parameter space (see Appendix~\ref{app:lps-filters}). For all seven datasets, the CNN's increase in accuracy over the f-ERM is too small to be of any practical significance. Like most contemporary quantum computing systems, IBM Q processors experience a mixture of coherent and stochastic errors, and the prediction accuracy of these CNNs is between what we observed in simulations with purely stochastic errors (Sections~\ref{sec:5Q-sims} and~\ref{sec:49Q-sims}) and purely coherent errors (Section~\ref{ssec:coherent}). We therefore conjecture that a neural network method that can accurately predict $s(c)$ when coherent errors are a significant proportion of the total error budget is the primary advance required to obtain useful neural network models for $s(c)$.

Real quantum processors are not stable, i.e., their error processes vary with time \cite{Mavadia2018-ki, Proctor2020-iz, Huo2018-ej}, which implies that a processor's capability function $s(c)$ depends on time $t$ (and possibly other context variables---see Section~\ref{sec:capability}). Because our datasets do not include execution time in the data encoding, it is infeasible to learn the impact of any processes that vary over time scales that are longer than the time for running individual circuits (processes that varying within the execution time of a single circuit can be learned, as demonstrated above). Therefore the magnitude of a processor's instability provides a lower bound on how accurate a model trained on that data can be. The magnitude of a processor's instability can be quantified by using the estimate of $s(c)$ for each circuit from the second pass through all the circuits---denoted by $\hat{s}'(c)$---as a prediction for $s(c)$ on the test data (obtained in pass 1). This prediction for $s(c)$, which we call the \emph{stability baseline model} (SBM), is shown in the fourth row of Fig.~\ref{fig:experimental-combined}. We observe that the CNNs (and the f-ERMs) do not achieve the prediction accuracy of the SBMs. However, for all but two processors the average $L^1$ error of the CNN is less than two times the average $L^1$ error of the SBM (see Fig.~\ref{fig:experiment-metrics}).

\section{Conclusions}\label{sec:discussion}
Understanding a quantum processor's computational power requires knowledge of what quantum circuits it can run with low probability of error, and this can be formalized using the concept of a \emph{capability function} $s(c)$. But modelling a quantum processor's capability $s(c)$ is hard, as $s(c)$ depends on a processor's unknown errors and how those errors interact with each circuit $c$. Parameterized error models can be used to predict $s(c)$, but simple, scalable error models fail to accurately model $s(c)$. More complex error models like process matrices are not scalable---yet they still often fail to model $s(c)$ accurately \cite{Blume-Kohout2020-fl}. In this work we have investigated using neural networks to learn an approximation to $s(c)$. We have formalised this prediction problem [i.e., defining $s(c)$] and outlined a very general method for modelling $s(c)$ with neural networks when training data can be gathered efficiently (e.g., for process fidelity and probability of successful trial).

We have presented a CNN architecture and circuit encoding for modelling $s(c)$, and we have shown that these CNNs can accurately model $s(c)$ in the presence of Markovian or non-Markovian local Pauli stochastic noise. Interestingly, we have demonstrated that a CNN vastly outperforms an ERM (error rates model) on some simple non-Markovian error models, demonstrating the power of neural networks to model the impact of errors that are outside the standard framework of parameterized error models. However, a practical technique for modelling $s(c)$ must be able to learn a good approximation to $s(c)$ in the presence of all the kinds of error that real processors commonly experience, and this is not yet the case with our CNNs. We find that they less accurately model $s(c)$ when coherent errors are dominant, which limits the applicability of these networks to real systems. A promising approach to solving this problem is to use a circuit encoding (and a corresponding network architecture) that contains detailed error sensitivity information, generalizing the error sensitivity channels in our encoding. Our error sensitivity channels encode a circuit's sensitivity to local Pauli stochastic errors, and they help our CNNs perform excellently on local Pauli stochastic error models. However, whether it is possible to efficiently summarize general error sensitivity information for general circuits is an open problem.

The CNN architecture we have employed is well-suited to our prediction problem in a number of ways, e.g., a convolutional filter jointly analyzes information from a temporal neighbourhood of a gate. But this architecture also has some features that likely limit its ability to accurately model $s(c)$. Firstly, spatial relationships between qubits are not explicitly encoded, yet it is likely to be of significance for predicting $s(c)$ due to the importance of effects like localized crosstalk errors. Second, each convolutional filter in a CNN is transitionally invariant, yet error processes in quantum computers are not: different qubits experience different errors and at different rates. So, although we have found that a CNN can accurately learn to predict $s(c)$ under error models that are not spatially invariant, it is not optimized for this task. We therefore conjecture that alternative neural network architectures---such as graph neural network---will substantially outperform CNNs on the capability learning problem. Indeed, a promising method for predicting $s(c)$ using a graph neural network was recently demonstrated by Wang \textit{et al.}~\cite{Wan22}. It is possible that a neural network and data encoding that combines error sensitivity information with graph neural networks---i.e., a physics-informed neural network model of $s(c)$---could learn accurate models for $s(c)$ on real quantum processors, enabling fast and accurate predictions of a processor's computational capabilities.

\bibliographystyle{unsrt}
\bibliography{ieee-biblio}

\appendix

\section{Previous Work}\label{app:prevwork}

In this appendix, we contextualize our work by summarizing other efforts to model a quantum processor's capability function using neural networks. First, we focus on attempts to model a circuit's \textit{probability of successful trial}, which can be thought of as a kind of fidelity estimation. Probability of successful trial (PST) is a generalization of a definite outcome circuit's success probability to arbitrary circuits. For an arbitrary quantum circuit $c$, $\textrm{PST}(c)$ is defined as the success probability of the definite outcome circuit created by concatenating $c$ with $c^{-1}$. PST is strongly correlated with state fidelity \cite{Wan22}, although  $\textrm{PST}(c)$ is not a reliable estimate of state fidelity (errors can echo away in Loschmidt echos). Two works have used neural networks to predict PST for arbitrary quantum circuits. Liu \textit{et al.}~\cite{Liu20} used shallow neural networks to predict the PST of both random and algorithmic circuits, while Wang \textit{et al.}~\cite{Wan22} used graph transformers to perform the same task. Both of these works are complementary to ours, as Refs.~\cite{Wan22, Liu20} use different neural network architectures.  A consequence of this is that Refs.~\cite{Wan22, Liu20} use circuit encodings that differ from our image-based encoding. Liu \textit{et al.}~represented each circuit as a tuple containing: (i) the width of the circuit; (2) the depth of the circuit; (3) the total number of each single-qubit and two-qubit gates in the circuit; (4) the number of measurements; and (5) a dictionary listing the target and control qubits for each two-qubit gate in the circuit. Wang \textit{et al.}~represented each circuit as a colored undirected graph, with colored vertices representing gates. Additional information, like one- and two-qubit gate error rates were also embedded in the graph. 

To our knowledge, this paper is the first to use CNNs to predict circuit success probability, but other papers have used CNNs for state fidelity estimation. Amer \textit{et al.}~\cite{Ame22} used one-dimensional CNNs to predict the state fidelity between the output state of shallow (3 to 5 layers) 1, 3, and 5-qubit circuits run on IBM Q devices. Vadali \textit{et al.}~\cite{Vad22} similarly focused on shallow circuits, but instead chose to predict state fidelity using a 3-dimensional CNNs. They also analyzed wider circuits (up to 25 qubits), which limited their efforts to working with simulated local depolarization and crosstalk errors.  In common with our work, Amer \textit{et al.}~and Vadali \textit{et al.}~use a modified version of one-hot encoding to encode circuits. However, the differences in the dimensions of the CNNs lead to different encoding schemes. Amer \textit{et. al}~flatten each circuit into a one-dimensional vector, while Vadali \textit{et. al}~maintain local connectivity information by assigning each gate a two-dimensional coordinate on a grid. 

Scalability is a significant problem when using neural network to model state fidelity. Amer \textit{et al.}~avoid this problem by focusing on short, few-qubit circuits, although it is arguable that useful neural network models for a processor's capability will need to be able to predict deep, many-qubit circuits. This allows  Amer \textit{et al.}~to perform state tomography to gather their training and test data (state tomography on states produced by general $n$-qubit circuits is inefficient in $n$). Vadali \emph{et al.}~instead focus on Clifford and Clifford-reducible circuits (two classes of efficiently classically simulable circuits) simulated under symmetric local depolarization noise with two-qubit gate $ZZ$-crosstalk. They then investigate how well their networks' extend to generic circuits simulated under the same noise model. Our work differs from that of Amer \textit{et al.}~and Vadali \emph{et al.}~in several important ways. In this work we examine how robust our CNNs are (e.g., how well does a CNN model generalize to out-of-distribution circuits?) as well as how a CNN's performance scales as a function of dataset size and quality. We also include error sensitivity information in the circuit encoding, we demonstrate the successful application of a neural network approach to modelling $s(c)$ with non-Markovian errors, and we show that CNNs fail to accurately model $s(c)$ in the presence of coherent noise.

\section{Parameterized error models}\label{app:max-markovian-model}
Conventional approaches to modelling a capability function $s(c)$ often involve constructing a parameterized error model. In this appendix we review the maximal Markovian error model. The most widely-used models for a quantum computer's errors---including the ERMs used in this work---can be constructed by placing constraints on the maximal Markovian error model. In the maximal Markovian model each imperfect $n$-qubit layer of gates $l \in \mathbb{L}_n$ is modelled by a fixed but unknown CPTP map on $n$-qubits, a state preparation by a $n$-qubit density matrix $\rho$, and a measurement by an $n$-qubit positive operator-valued measurement $M$. Each $n$-qubit CPTP map has $4^n(4^n-1)$ parameters. So, the maximal Markovian model contains 
\begin{equation}
N_p(n) = 4^n(4^n-1)
N_L +N_{\textrm{SPAM}} = O(16^nN_L)
\end{equation}
parameters where $N_L$ is the number of possible circuit layers and $N_{\textrm{SPAM}}$ is the number of parameters in $\rho$ and $M$.

In principle, the parameters of the maximal Markovian model can learned from GST (up to a gauge freedom \cite{Nie21}), and the elements of the learned model can be composed to compute the model's prediction for $s(c)$ for any circuit $c$ \cite{Nie21}. But it is infeasible to estimate all $O(16^nN_L)$ of this model's parameters when $n \gg 1$. Tractable models with a polynomial number of parameters can be obtained by placing restrictions on the parameters of this model. This is the basis for the existing and nascent scalable models for a quantum computer's errors, including ``low-weight'' error models \cite{Blu22}, restricted Pauli stochastic models \cite{Harper2020-te}, crosstalk-free models \cite{Sarovar2020-pz}, and the ERMs we use in the main text.

\section{The motivation for success probability learning}\label{app:spl}
In this appendix we briefly expand on our explanation for why we focus on predicting the success probabilities of mirror circuits in this work. Methods that can accurately predict the success probabilities of mirror circuits are of little direct utility---no useful algorithms are mirror circuits, most quantum circuits are not definite outcome circuits, and we cannot \emph{a priori} assume that a neural network trained on mirror circuits will generalize to other families of circuit. We have chosen to consider the problem of modelling $s(c)$ for definite outcome circuits because it is a convenient setting in which to explore capability learning methods for the following two reasons. First, training data is easy to gather: estimating $s(c)$ with precision $\delta$ for a circuit $c$ simply requires sampling from that circuit's probability distribution (either in simulation, or on a real system) $O(1/\delta)$ times. In contrast, although mirror circuit fidelity estimation (MCFE) is efficient in $n$, MCFE requires running many circuits to estimate $s_F(c)$ \cite{Proctor2022-es}. Second, the problem of predicting circuit success probabilities retains many of the aspects of fidelity learning, and we conjecture that a neural network method that can accurately model $s(c)$ when trained on circuit success probabilities will, with only minor adaptions, be able to accurately model $s_F(c)$ when trained on circuit process fidelities. One reason for this conjecture is that MCFE estimates a circuit $c$'s process fidelity by simple data processing on estimations of the success probabilities from a set of mirror circuits based on $c$.

\section{Detailed introduction to CNNs}\label{sec:cnns:app}
Our aim is to learn an approximation to the mapping from circuits $c$, encoded into $n \times d \times h$ images $I(c)$, to $s(c) \in [0, 1]$. Therefore, our CNNs are functions 
\begin{equation}\label{eqn:CNN}
\text{CNN}: \mathbb{R}^{n \times d\times h}\rightarrow [0,1].
\end{equation}
Our CNNs are built out of three kinds of layers: convolutional layers (\textbf{conv}), pooling layers (\textbf{pool}), and dense layers (\textbf{dense}). These CNNs consist of interleaved convolutional and pooling layers, followed by a sequence of dense layers, as illustrated in Fig.~\ref{fig:nm-architecture}.

Convolutional layers create \emph{feature maps} by convolving an input image with learnable kernels. A convolutional layer for an input image $I^{(\textrm{in})} \in \mathbb{R}^{n_{\textrm{in}} \times d_{\textrm{in}} \times h_{\textrm{in}}}$ is specified by an activation function $f$, a kernel shape $(k_{w}, k_{d})$, and $h_{\textrm{out}}$ convolutional filters $ \mathbb{R}^{n_{\textrm{in}} \times d_{\textrm{in}} \times h_{\textrm{in}}} \to  \mathbb{R}^{n_{\textrm{in}} \times d_{\textrm{in}}}$ containing learnable parameters. The convolutional layer is a map
\begin{equation}
    \textbf{conv}: \mathbb{R}^{n_{\textrm{in}} \times d_{\textrm{in}} \times h_{\textrm{in}}} \rightarrow \mathbb{R}^{n_{\textrm{in}} \times d_{\textrm{in}} \times h_{\textrm{out}}}
\end{equation}
constructed by ``stacking'' the output of the $h_{\textrm{out}}$ filters. Each convolutional filter is defined by a learnable kernel $K^{(h)} \in\mathbb{R}^{k_{w} \times k_{d}\times h_{\textrm{in}}}$ and a learnable bias $b_h \in \mathbb{R}$. The filter turns the three-dimensional input image $I^{(\textrm{in})}$ into a two-dimensional feature map, by convolving the kernel with $I^{(\textrm{in})}$---i.e., by sliding the kernel across the image and, at each location, taking the inner product of the kernel with the local image patch---then adding the bias ($b_h$) to each pixel in the resultant two-dimensional image, and finally applying the activation function ($f$). So the three-dimensional image output by the convolutional layer [$I^{(\textrm{out})}$] is given by
\begin{equation}\label{eqn:conv}
    I_{ijh}^{(\textrm{out})} = f\left(\sum_{i'=-\lfloor \nicefrac{k_{w}}{2} \rceil}^{\lceil \nicefrac{k_{w}}{2} \rceil}\sum_{j'=-\lfloor \nicefrac{k_{d}}{2}  \rfloor}^{\lceil \nicefrac{k_{d}}{2}  \rceil} \sum_{h'=1}^{h_{\textrm{in}}} K_{i'j'h'}^{(h)}I^{(\textrm{in})}_{(i+i'),(j+j'),h'} + b_h\right),
\end{equation}
for $i=1,\dots, n_{\textrm{in}}$, $j=1,\dots, d_{\textrm{in}}$, and $h=1,\dots,h_{\textrm{out}}$. Note the edges of the input image are padded with zeros, e.g., by definition $I^{(\textrm{in})}_{-1jh} = 0$. In our work, the activation function $f:\mathbb{R}\rightarrow \mathbb{R}$ is the rectified linear unit (ReLU):
\begin{equation}
    \text{ReLU}(x) = \max(0,x).
\end{equation}

Convolutional layers are useful for extracting predictive features from quantum circuits $I(c)$ because they can create feature maps that identify the locations (and the number of instances) of specific patterns of gates in $c$. This is relevant to predicting $s(c)$ because particular patterns of gates can increase the failure rate of circuits \cite{Sarovar2020-pz, Pro22}. For example, error rates can increase when a particular gate is repeated multiple times in a row (known as serial context dependence) or when certain gates are applied in parallel (known as parallel context dependence or crosstalk) \cite{Sarovar2020-pz}. Convolutional filters exist that, e.g., find all instances of sequential applications of the same gate in $I(c)$ (see Section~\ref{ssec:double-trouble} and Appendix~\ref{app:filters}). Furthermore, in Appendix~\ref{app:lps-filters} we present convolutional filters that extract feature maps from our circuit encoding that are sufficient to approximate $s(c)$ under a local stochastic Pauli error model (this family of parameterized models is defined in Section~\ref{sec:49Q-sims}), and local stochastic Pauli errors constitute a significant proportion of the total error in many systems.

In our networks, each convolutional layer is followed by a pooling layer (see Fig. \ref{fig:nm-architecture})
\begin{equation}
    \textbf{pool}: \mathbb{R}^{n_{\textrm{in}}\times d_{\textrm{in}}\times h_{\textrm{in}}}\rightarrow\mathbb{R}^{n_{\textrm{out}}\times m_{\textrm{out}}\times h_{\textrm{out}}}, 
\end{equation}
where $n_{\textrm{out}}\leq n_{\textrm{in}}$ and $d_{\textrm{out}}\leq d_{\textrm{out}}$. Pooling layers reduce the size of an image, by partitioning the image into distinct $(p_{w},p_{d})$-shaped segments and, in each channel $h$, replacing each such segment with the maximum or average value within that segment. Pooling layers, which contain no learnable parameters, are included in the networks for dimensionality reduction \footnote{Reducing the size of the image reduces the size of the input to the first dense layer, therefore reducing the number of weights that must be learned for that layer.}.

The dense layers of a CNN (see Fig.~\ref{fig:nm-architecture}) are used to process the final feature maps in order to make a prediction. Each dense layer is a map
\begin{equation}
    \textbf{dense}: \mathbb{R}^{n_{\textrm{in}}}\rightarrow \mathbb{R}^{n_{\textrm{out}}}
\end{equation}
that consists of $n_{\textrm{out}}$ artificial neurons $\lbrace u_i\rbrace_{i=1}^{n_{\textrm{out}}}$ and a non-linear activation function $f:\mathbb{R}\rightarrow\mathbb{R}$ (we again used ReLU, except for the final layer). Each neuron $u :\mathbb{R}^{n_{\textrm{in}}} \to \mathbb{R}$ is defined by a learnable weight vector, $w\in\mathbb{R}^{n_{\textrm{in}}}$, and a learnable bias, $b$. The mapping defined by a neuron is $u(v) = f(w^Tv+b)$, and the output of the layer is $v'=(u_1(v),\dots,u_{n_{\textrm{out}}}(v))^T$. Our networks terminate with a dense layer containing a single neuron with a sigmoid activation function. This guarantees that our network's output is in $[0,1]$, and thus represents a probability.

\section{CNNs can approximate local stochastic Pauli error models}\label{app:lps-filters}
In this appendix we provide CNN filters that approximate a local stochastic Pauli errors model, under the assumption of no parameterized gates. This is a proof that such a network exists. We are not claiming that a network with this structure and these weights is learned when we train a CNN on data from a local Pauli stochastic error model (and indeed it is not). For a stochastic Pauli errors model,
\begin{equation}
s(c)\approx \prod I(c)_{ijh_P} (1-\epsilon_P(c_{ij})). \label{eq:lp-s}
\end{equation}
This approximate formula for $s(c)$ accounts for bias in the errors (i.e., that a particular Pauli operator does not change the state of the qubits if it is applied to a state that is an eigenstate of that Pauli operator), by making using of the error sensitivity channels. The approximation used here is that two Pauli errors, that occur at different circuit locations, never cancel, which is a very good approximation in random circuits when $n\gg 1$ \cite{Polloreno2023-xa}.

To demonstrate that we can approximately reproduce Eq.~\eqref{eq:lp-s} using a CNN we will specify a convolution filter for every possible gate and qubit pair and every possible $P$. Assume that $I(c)$ has shape $w \times d \times \text{n}_{\text{channels}}$. For a gate $G$ (encoded by a 1 in channel $\text{ch}(G)$) qubit $i$ and Pauli $P$ we define the following filter $K$, which we use with a bias of $-1$: $K$ is shape $w \times 1 \times \text{n}_{\text{channels}} $ and it contains zeros everywhere except on qubit $i$. For that qubit, it contains a 1 in the error sensitivity channel for $P$ and $\epsilon_P(G,i)$ in the channel for the gate $G$. Consider the single convolutional layer that is a collection of these kernels. This produces an image whereby the elements are all zero except for $d\times w$ elements which are the $\epsilon_P(c_{ij})$. We can then approximate $s(c)$ simply by the sum of all the elements of this tensor. To obtain the (better) approximation to $s(c)$ stated in Eq.~\eqref{eq:lp-s} we need to take the product of $1-\epsilon_P(c_{ij})$---and a dense network can be trained to approximate the product of one minus each of its input.

\section{Markovian error models and simulations}\label{app:noise-models}
This appendix contains additional details on how we constructed each of the three simulated Markovian error models used in this paper. In order, they are: (i) the 5-qubit Markovian local Pauli stochastic error model from Section~\ref{sec:5Q-sims}; (ii) the 5-qubit Markovian local Hamiltonian error model used in Section~\ref{ssec:coherent}; and (iii) the 49-qubit base Markovian local Pauli stochastic error model from Section~\ref{sec:49Q-sims}. We first explain how the 5-qubit Markovian local Pauli stochastic error model was constructed, before enumerating the alternations made to construct the Markovian local Hamiltonian model. We then conclude with a brief description of the 49-qubit base Markovian local Pauli stochastic error model.

The 5-qubit local Pauli stochastic error model was specified using the error generator formalism of Ref.~\cite{Blu22}. It consists of operation-dependent errors randomly sampled according to a two-step process [i.e., there are independently sampled error rates for each gate and qubit(s) pair]. First, an error rate was selected for each type of one- or two-qubit gate by uniformly sampling a value in $[0,1]$ and multiplying it by a pre-determined maximum error rate. Then a single one-qubit (resp. two-qubit) stochastic error generator was uniformly sampled for each one-qubit (resp. two-qubit) gate. Each randomly sampled error generator was then assigned its gate's error rate. Maximum one- and two-qubit gate error rates of .25\% and 1\% were chosen for the Pauli stochastic model. The exact error generators and rates are available in the Supplementary Materials.

The 5-qubit local Hamiltonian error model (see Section~\ref{ssec:coherent}) was also specified using the error generator formalism. Operation-dependent Hamiltonian error generators were uniformly sampled. Each randomly sampled error generator was assigned an error rate of 5\%. The sampled error generators are available in the Supplementary Materials.

The 49-qubit base Markovian local Pauli stochastic error model (see Section~\ref{sec:49Q-sims}) was specified by directly sampling the rates of Pauli $X$, $Y$ and $Z$ error rates (this differs slightly from the parameters in the error generator formalism) for each gate and each qubit on which that gate acts. Unlike for the 5-qubit local Pauli stocahstic error model, we did not use maximally biased errors, so each single-qubit gate and qubit pair is assigned three error rates (and each two-qubit gate and qubits pair is assigned six error rates). These error rates were uniformly sampled from $[0,.0001]$. Readout error rates of $.0001$ were also assigned to each qubit.

\section{Convolutional filters for Double Trouble}\label{app:filters}
In this appendix we state a set of four convolutional filters that enable the identification of all instances of sequential CNOT gates in a circuit. Consider a circuit $c$ and assume that there is a CNOT gate in layer $j$ that acts on qubit $i$ (and some other qubit $i'$). Then, in our tensor encoding $I(c)$ of the circuit, $I(c)_{ijk} = \pm 1$ where $k$ is one of the four channels used to encode CNOT gates. Therefore, there is also a CNOT gate on qubit $i$ in the layer before $j$ (i.e., $j-1$) if and only 
\begin{equation}
\sum_{k \in \mathbb{C}_{\textsc{cnot}}} |I(c)_{i(j-1)k}| + |I(c)_{ijk}| = 2,
\end{equation}
where the summation is over the four CNOT channels. There is a set of four $1 \times 2 \times n_{\textrm{channels}}$ convolutional kernels $K$ and biases $b$ whereby one of these four convolutional filters outputs a non-zero pixel if and only if this criteria is satisfied. These four filters correspond to $K_{1,1,k} = \pm\nicefrac{1}{2}$ and $K_{1,2,k} = \pm\nicefrac{1}{2}$ if $k$ is a CNOT channel, with $K_{1,1,k} = K_{1,2,k} = 0$ otherwise, all with a bias of $b = -\nicefrac{1}{2}$. The output $p(i,j)$ when a  $1 \times 2 \times n_{\textrm{channels}}$ kernel is applied at location $(i,j)$ is
\begin{equation}
   p(i,j) = \textrm{ReLU}\left(\sum_{j'=0}^{1}\sum_{k} I(c)_{i(j-j')k}K_{i(j-j')k} - b\right),
\end{equation}
where $\textrm{ReLU}$ is the ReLU activation function. Therefore if, e.g., $K_{1,1,k} = K_{1,2,k} = \nicefrac{1}{2}$ then $p(i,j) = \nicefrac{1}{2}$ if $\sum_{k \in \mathbb{C}_{cnot}}[I(c)_{i(j-1)k} + I(c)_{ijk}]= 2$ and otherwise $p(i,j) = 0$.

\section{Additional Experimental Data}\label{app:experimental-predictions}
\begin{figure*}
\includegraphics[width=16cm]{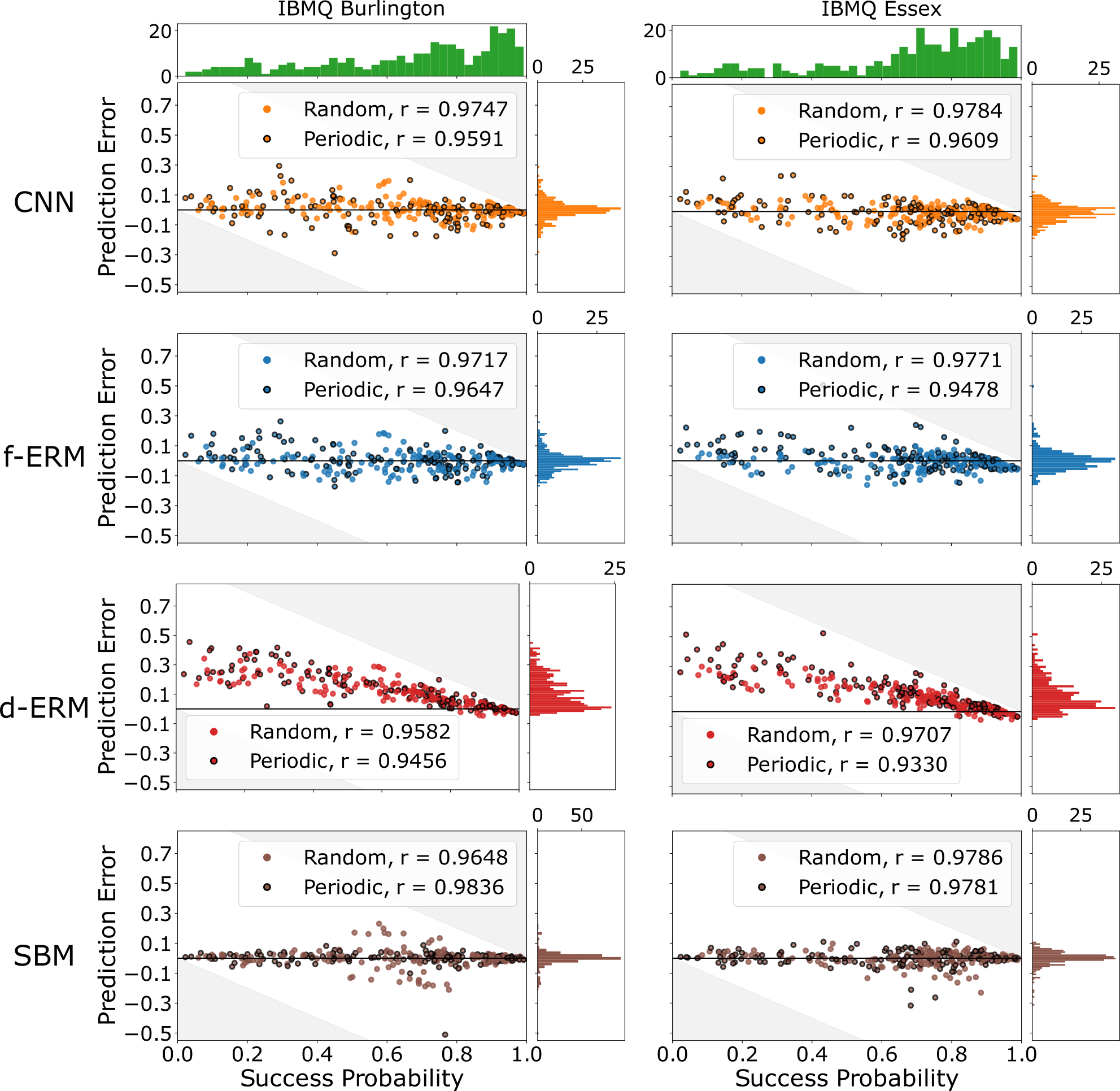} 
\caption{\textbf{Predicting the capabilities of cloud-access IBM Q processors (continued)}. Plots of the prediction errors for two of the four processors not shown in Fig.~\ref{fig:experimental-combined} of the main text. See the caption of Fig.~\ref{fig:experimental-combined} for details.}
\label{fig:app-experimental-predictions-1}
\end{figure*}

\begin{figure*}
\includegraphics[width=16cm]{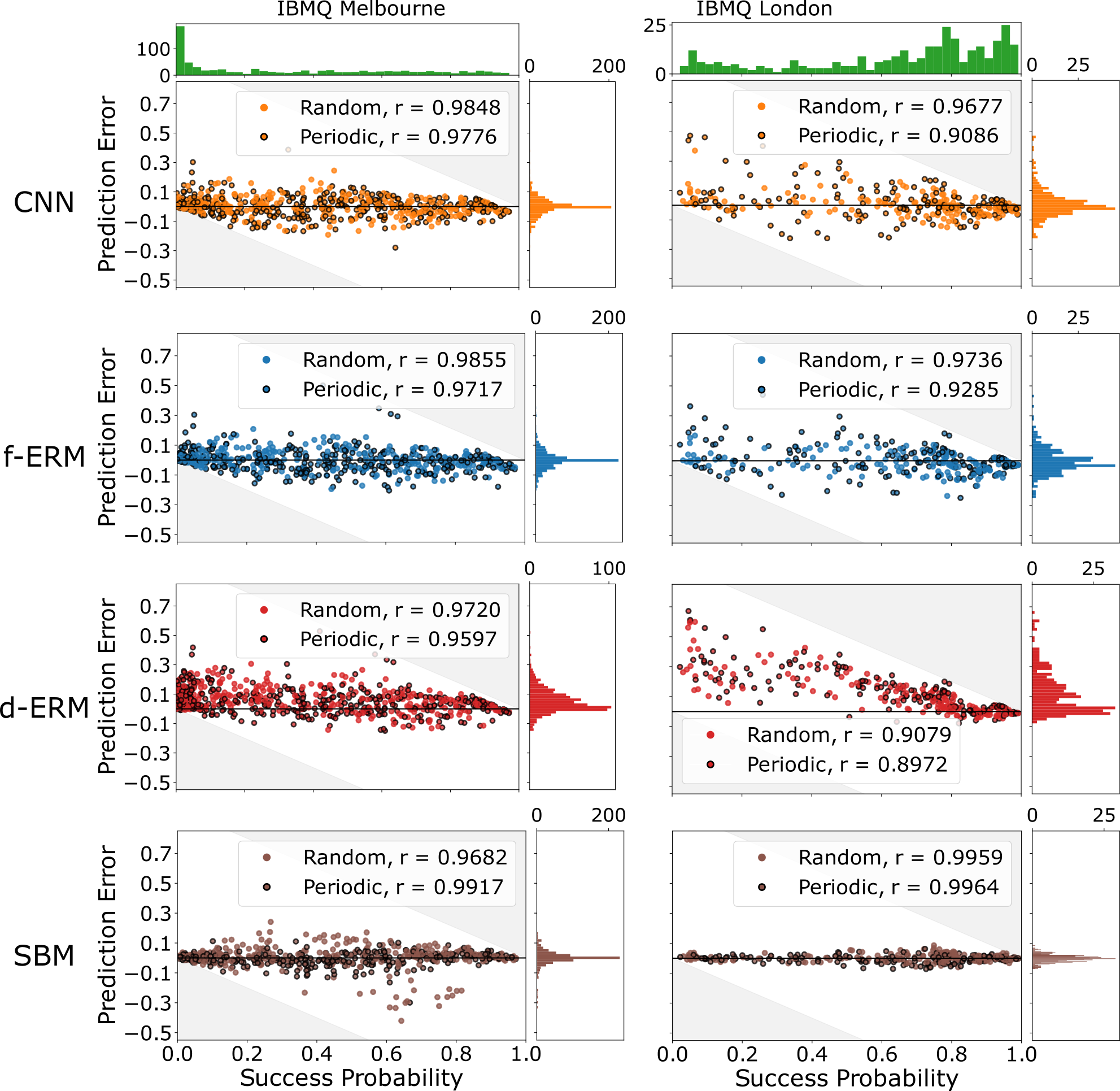} 
\caption{\textbf{Predicting the capabilities of cloud-access IBM Q processors (continued)}. Plots of the prediction errors for two of the four processors not shown in Fig.~\ref{fig:experimental-combined} of the main text. See the caption of Fig.~\ref{fig:experimental-combined} for details.}
\label{fig:app-experimental-predictions-2}
\end{figure*}

In the main text we presented plots of the prediction error for three of the seven IBM Q processors (see Fig.~\ref{fig:experimental-combined}). In Figs.~\ref{fig:app-experimental-predictions-1} and~\ref{fig:app-experimental-predictions-2} of this appendix we present equivalent plots for the remaining four processors.

\section*{Acknowledgements}
The authors would like to thank Brent Haglund for help designing the paper's figures. Sandia National Laboratories is a multi-mission laboratory managed and operated by National Technology \& Engineering Solutions of Sandia, LLC (NTESS), a wholly owned subsidiary of Honeywell International Inc., for the U.S. Department of Energy’s National Nuclear Security Administration (DOE/NNSA) under contract DE-NA0003525. This written work is authored by an employee of NTESS. The employee, not NTESS, owns the right, title and interest in and to the written work and is responsible for its contents. Any subjective views or opinions that might be expressed in the written work do not necessarily represent the views of the U.S. Government. The publisher acknowledges that the U.S. Government retains a non-exclusive, paid-up, irrevocable, world-wide license to publish or reproduce the published form of this written work or allow others to do so, for U.S. Government purposes. The DOE will provide public access to results of federally sponsored research in accordance with the DOE Public Access Plan. We acknowledge the use of IBM Quantum services for this work. The views expressed are those of the authors, and do not reflect the official policy or position of IBM or the IBM Quantum team.

\section*{Data availability}
Data, code, and models are available at Ref.~\cite{dataset}. All circuit sampling and error model simulations were performed using \texttt{pyGSTi} \cite{Nielsen2020-rd} or custom code that can be found at Ref.~\cite{dataset}. All neural networks were constructed and trained using \texttt{TensorFlow} with \texttt{Keras} \cite{ten15, ker15}.

\EOD

\end{document}